\documentclass{article}
\usepackage[affil-it]{authblk}
\usepackage{graphicx}
\usepackage[space]{grffile}
\usepackage{latexsym}
\usepackage{textcomp}
\usepackage{longtable}
\usepackage{appendix}
\usepackage{multirow,booktabs}
\usepackage{amsfonts,amsmath,amssymb}
\newif\iflatexml\latexmlfalse
\usepackage[utf8]{inputenc}
\usepackage{subfigure}

\usepackage[stable]{footmisc}
\usepackage{mathrsfs}
\usepackage[]{natbib}
\bibpunct{(}{)}{;}{a}{}{,} 
\usepackage{txfonts}
\usepackage{xspace}
\usepackage{ifthen}
\usepackage[tables,usenames,dvipsnames]{xcolor}
\usepackage[breaklinks, colorlinks, citecolor=blue]{hyperref}
\usepackage[english]{babel}
\usepackage{fullpage}
\usepackage{listings}
\definecolor{light-gray}{gray}{0.95}
\usepackage{aas_macros}
\newcommand{\mksym}[1]{\ifmmode {\rm #1}\else #1\fi}

\newcommand{\eV}{\,\text{eV}}

\providecommand{\omb}{\omega_{\mathrm{b}}}
\providecommand{\omc}{\omega_{\mathrm{c}}}

\newcommand{\nnu}{N_{\rm eff}}
\newcommand{\neff}{\nnu}

\newcommand{\As}{A_{\rm s}}

\newcommand{\ns}{n_{\rm s}}

\providecommand{\CAMB}{{\tt camb}}
\providecommand{\COSMOMC}{{\tt CosmoMC}}
\providecommand{\CLASS}{{\tt class}}
\providecommand{\LCDM}{{$\rm{\Lambda CDM}$}}

\providecommand{\minuit}{{\tt MINUIT}\xspace}

\providecommand{\CAMEL}{{\tt CAMEL}}
\newcommand{\onesig}[1]{(68\%, \text{#1})}

\providecommand{\sorthelp}[1]{}

\newcommand{\wwwcamb}{\url{http://camb.info}}
\newcommand{\wwwclass}{\url{http://class-code.net/}}
\newcommand{\wwwminuit}{\url{http://seal.web.cern.ch/seal/work-packages/mathlibs/minuit/index.html}}

\providecommand{\Planck}{\textit{Planck}}
\providecommand{\planck}{\Planck\xspace}

\newcommand{\class}{\CLASS\xspace}
\newcommand{\lcdm}{\LCDM\xspace}
\newcommand{\camel}{\CAMEL\xspace}

\newcommand{\highell}{\ifmmode {\rm high}-\ell \else high-$\ell$\fi}
\newcommand{\hiell}{\ifmmode {\rm high}-\ell \else high-$\ell$\fi}
\newcommand{\lowell}{\ifmmode {\rm low}-\ell \else low-$\ell$\fi}

\newcommand{\hlp}{\texttt{Hillipop}}

\newcommand{\cl}{\ensuremath{C_\ell}}
\newcommand{\lik}{\ensuremath{\mathscr{L}}}
\newcommand{\lnAs}{\ln( 10^{10} \As)}
\newcommand{\bflike}{\texttt{bflike}}

\newcommand{\chimin}{\ensuremath{\chi^2_\textrm{min}}}
\newcommand{\dchi}{\ensuremath{\Delta \chi^2}}

\newcommand{\parfile}{\texttt{hlpTT\_bflike\_LCDM.par}\xspace}

\begin{document}

\title{Agnostic cosmology in the \CAMEL\ framework}

 \author{S. Henrot-Versill\'e, O. Perdereau, S. Plaszczynski \footnote{corresponding author \url{plaszczy@lal.in2p3.fr}}, B. Rouill\'e d'Orfeuil, M. Spinelli and M. Tristram.\\

\textit{Laboratoire de l'Acc\'el\'erateur Lin\'eaire, Univ. Paris-Sud, CNRS/IN2P3, Universit\'e Paris-Saclay, Orsay, France.} 
}

\date{\today}

\maketitle

\begin{abstract}

Cosmological parameter estimation is traditionally performed in the Bayesian context. 
By adopting an "agnostic" statistical point of view, we show the interest of confronting the Bayesian results to a frequentist approach based on profile-likelihoods.

To this purpose, we have developed the \textit{Cosmological Analysis with a Minuit Exploration of the Likelihood} (\CAMEL) software.  Written from scratch in pure \texttt{C++}, emphasis was put in building a clean and carefully-designed project where new data and/or cosmological computations can be easily included. 

CAMEL incorporates the latest cosmological likelihoods and gives access \textit{from the very same input file} to several estimation methods:
\begin{itemize}
\item A high quality Maximum Likelihood Estimate (a.k.a "best fit") using \minuit,
\item profile likelihoods,
\item a new implementation of an Adaptive Metropolis MCMC algorithm that relieves the burden of reconstructing the proposal distribution.
\end{itemize}

We present here those various statistical techniques and 
roll out a full use-case that can then used as a tutorial. 
We revisit the \lcdm\ parameters determination with the latest \planck\ data and give results with both methodologies.
Furthermore, by comparing the Bayesian and frequentist approaches, we discuss a "likelihood volume effect" that affects  
the optical reionization depth when analyzing the high multipoles part of the \planck\ data. 

The software, used in several \planck\ data analyzes, is available from \url{http://camel.in2p3.fr}. Using it does not require advanced {\tt C++} skills.%
\end{abstract}%

\begin{figure}[h!]
\begin{center}
\includegraphics[width=0.25\columnwidth]{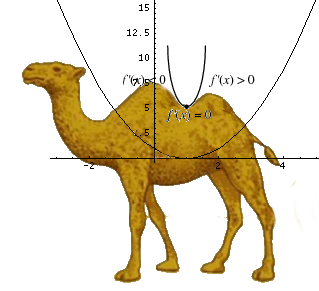}
\end{center}
\end{figure}

\pagebreak
\tableofcontents
\pagebreak

\section{Introduction}
\label{sec:Introduction}

Since the 2000's \citep{Christensen_2001}, the adoption of Monte Carlo Markov Chain (MCMC) techniques by the cosmological community has promoted the acceptance of Bayesian methodology for parameter estimation. The possible (although inefficient) sampling of a high dimensional space allows, invoking Bayes theorem, to reconstruct posterior distributions assuming some prior degree of belief (rarely discussed). MCMC usage was popularized in cosmology by the \COSMOMC\ package \citep{cosmomc} which uses an optimized version of the Metroplis-Hastings algorithm (see also \texttt{MontePython} \citep{montepython}). There exist today several implementations of other algorithms, as \texttt{MultiNest} \citep{multinest} or \texttt{PolyChord} \citep{Handley_2015} for Nested Sampling \citep{skilling2004}, or \texttt{emcee} \citep{emcee} for the affine invariant ensemble sampler \citep{goodman}. Several of them are packaged within the \texttt{CosmosSIS} package \citep{cosmosis} and some of them were compared in \citet{Allison_2014}. They are essentially written in \texttt{Fortran}/\texttt{python} and/or wrapped into multiple languages. All of them adopt the Bayesian paradigm (see \citet{BayesCosmo} for a review).

In other communities, as High Energy Physics, the frequentist approach is more traditional.
The reason for its absence in cosmology maybe lies in the difficulty of building precise profile-likelihoods because of the numerical noise inherent to the Boltzmann solver computations (Sect.\ref{sec:precision}). It was however shown to be feasible on real \planck\ data using an accurate minimizer and tuning the precision of the Boltzmann computations \citep{prof14}.

We wish to avoid the ideological debate about "who is right", and focus on the interest of confronting both methodologies. It is the goal of the \camel package to provide tools to compare the Bayesian and frequentist analyzes 
in order to better understand for instance the Bayesian priors effects or the shape of the likelihood.
We choose to write it in pure \texttt{C++} since we consider it as a proper language to develop a robust, clean and long term project. By \texttt{C++} we mean an object-oriented code with some level of abstraction and well defined design patterns in order to build a clean interface that anticipates future developments.
Fortunately users do not need to know what is under the hood  and one can plug in easily any new Boltzmann solver and/or likelihood.

Certainly the best way to present \CAMEL\ is by working out a full use-case. We will revisit the \lcdm\ cosmological parameters estimation with \planck\ data using the \hlp\ likelihood \citep{Like15} and compare the Bayesian and frequentist results. 
Not only will we show how to obtain precise best-fits and profile-likelihoods, but also propose a new implementation of an Adaptive scheme for the Metropolis MCMC algorithm that relieves the pain of reconstructing the proposal.
Sect. \ref{sec:pipeline} is a quick overview of the building blocks of parameter estimation.
Then Sect. \ref{sec:use_case} explains in minute-details how to obtain best-fits (Sect.\ref{sec:MLE}), profile-likelihoods (Sect.\ref{sec:profile}) and produce MCMC with the new Adaptive algorithm (Sect. \ref{sec:MCMC}). For each method, the basics will be reviewed.
This part concretely shows how to produce all the results and can therefore be used as a tutorial. 
Sect \ref{sec:comp} finally shows the interest of comparing both methods focusing on the study of the likelihood "volume effects".
These results, partially available in \citet{prof14} and \citet{cosmo15}, were however never compiled and detailed. The results on the volume effects that affect \planck's \hiell\ likelihoods are new and complement 
\citet{papierAlens}.

\section{The cosmological parameters estimation (CPE) pipeline}
\label{sec:pipeline}
\begin{figure}[h!]
\begin{center}
\includegraphics[width=0.7\columnwidth]{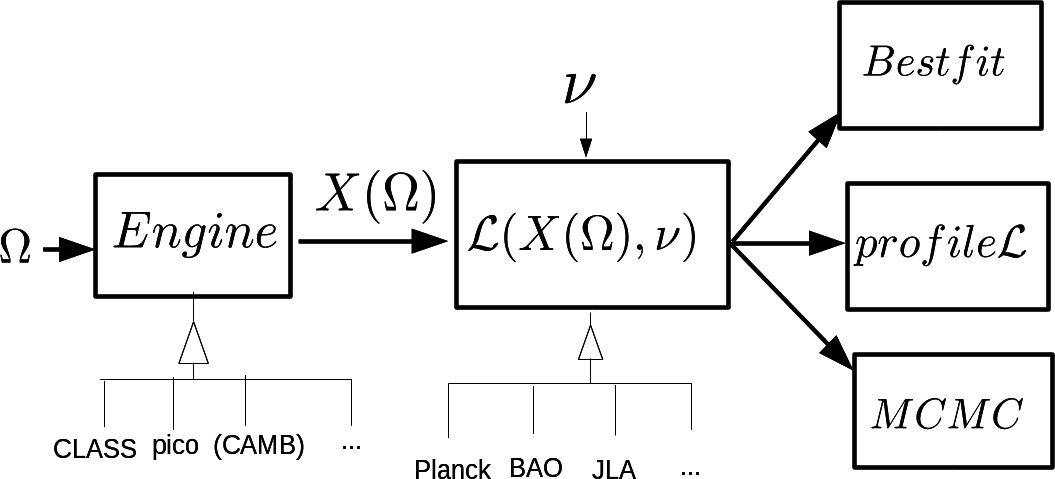}
\caption{{\label{fig:pipeline}
Schematics of the CPE pipeline. From left to right: $\Omega$ denotes the cosmological parameters that feed the Engine which computes some observable(s): $X(\Omega)$. $\nu$ denotes the nuisance parameters of the likelihood (or any external data necessary to its computation). The last 3 vertical boxes show different statistical methods.
Some implementations of engines and likelihoods available in \camel are derived from the base abstract classes.%
}}
\end{center}
\end{figure}

A CPE analysis can be factorized into 3 distinct pieces (see Fig. \ref{fig:pipeline}):
\begin{enumerate}
\item A code to compute some observables (as a set of \cl\ values for the CMB or $P(k,z)$ for the matter power spectra) given a set of cosmological parameters (called $\Omega$)  within a theoretical framework: $X(\Omega)$. In cosmology, the numerical codes that solve the background+perturbation equations are often called "Boltzman solvers". In \CAMEL\ such a tool is called an \textbf{Engine}. Today, \CLASS~({\wwwclass}) and \CAMB~({\wwwcamb}) are widely used but any analytic parametrization of some cosmological observable enters this category too.
\item The experimental data, which in their most complete statistical form are given in the form of a \textbf{likelihood}, i.e a function of $X$ + generally some extra nuisance parameters (called $\nu$). There is an implicit conversion here to cosmological parameters through the Engine: $\lik(\Omega,\nu)=\lik(X(\Omega),\nu)$.
In the following we will also  often use the $\chi^2$ function which is by definition $\chi^2\equiv-2\ln\lik$.
\item One then wants to extract some information about the parameters themselves (including the nuisance ones), so you need some \textbf{statistical methodology}. In the realm of \textit{estimators}, the Maximum Likelihood Estimator (MLE) is the King since it enjoys several nice properties (as being unbiased and of minimal variance asymptotically). It is also referred to as a "best fit" meaning the set of $(\hat \Omega,\hat \nu)$ values that gives the maximal likelihood value (or minimum $\chi^2$ one). This is a \textit{point estimate} and is a classical statistical notion. Since it is not very practical to realize what happens in a multi-dimensional space, we focus on individual (or a pair of) parameters : this is the domain of \textit{interval estimation}. Bayesian's will extract the p.d.f of the true parameters (posterior distribution), while frequentists will make statements on data covering the true (but unknown) values through the use of profile-likelihoods.
\end{enumerate}

\camel implements these 3 aspects through the abstraction mechanism allowed by Object-Oriented languages.
Currently the following concrete implementations (all in \texttt{C/C++}, including the likelihoods) are provided:
\begin{enumerate}
\item Engines: 
\begin{itemize}
\item \class \citep{classII}
\item \texttt{pico} \citep{Fendt_2007} 
\end{itemize}
\item Likelihoods:
\begin{itemize}
\item CMB: \planck \citep{Like15} \textit{Hillipop} \citep{Tristram_2005,Like15}, {\it SPT} \citep{Story_2013,Reichardt_2012,George_2015}, {\it ACT} \cite{Das_2014a},
\item CMB-lensing: \planck\ \citep{planck_lensing15},
\item BAO: {\it 6dF} \citep{Beutler_2011}, {\it WiggleZ} \citep{Blake_2011}, {\it SDSS} \citep{Padmanabhan_2012,Anderson_2012,Anderson_2014}: 1D and anisotropic,
\item SN1a : {\it JLA} \citep{Betoule_2014a}.
\end{itemize}
\item Statistics: best-fit, profile-likelihood, MCMC.
\end{enumerate}
See the Appendix for more details.

\section{A full \camel use-case}
\label{sec:use_case}
\subsection{Building and validating the parameter file}

Estimating cosmological parameter (with a full Boltzman solver) is a CPU demanding task and can hardly be performed on your laptop: one needs to rely on a computer cluster.
But before sending a lot of jobs to your farm, you should \textit{define} and \textit{test} your \textbf{parameter file} which describes all the cosmological and nuisance parameters (varied or fixed) , data , Engine and statistical method.
All the tests -in this section- can be run interactively.
We choose as a benchmark the determination of \lcdm\ parameters \planck high+low $\ell$ likelihoods using the \class\ engine.

\subsubsection{Parameter file}
We write in this part a parameter file named \parfile.

First comes the choice of the engine:
\begin{lstlisting}[caption=Definition of the engine]
engine=class
\end{lstlisting}

Then comes the choice of the likelihoods.
Here we will use the \planck \hlp\ likelihood (for the \hiell\ part) and \texttt{bflike} for the \lowell\ one (both discussed in \citet{cosmo15}).

\begin{lstlisting}[caption=Defining the likelihoods]
############################################################################
#hiell
HiLLiPOP=HiLLiPOP/DX11dHM_superExt_CO_TT.lik
#lowl
clikfile=planck_data/low_l/bflike/lowl_SMW_70_dx11d_2014_10_03_v5c_Ap.clik
##############################################################################
\end{lstlisting}

Then we define the parameters: 
the cosmological ones (i.e those entering the engine) must be stamped with the "cosm" keyword and follow the syntax of the Engine(here we follow the \class\ terminology). Nuisances, that depends on the likelihoods, should use the "nui" keyword.
Each parameter that  will be varied ("par") or not ("fix") should be given an initial value (no necessarily very precise), an expected standard deviation and some boundaries for finite exploration (which can be viewed as the flat prior borders for MCMC sampling or simply a box within which the MLE is searched).

\begin{lstlisting}[caption=Parameters definition for the \hlp+\bflike\ likelihoods]
##############PARAMETERS################################################
#cosmological parameters
#par/fix name            type     init         sigma     min     max
par   omega_b             cosm    0.02224     0.00027   0.017   0.027
par   omega_cdm           cosm    0.1192      0.0026    0.09    0.15
par   100*theta_s         cosm    1.0418      0.6E-04   1.03	1.05
par   tau_reio            cosm    0.07	      0.13E-01  0.01	0.20
par   log(10^10A_s)       cosm    3.07	      0.025     2.7     3.5
par   n_s                 cosm    0.96	      0.0070    0.9     1.1

#nuisances for hilipop
par   A_planck     	  nui	  1		0.001	   0.9 	  1.1
par   c0		  nui     0.		0.001     -0.05   0.05
par   c1		  nui     0.		0.001     -0.05   0.05
fix   c2		  nui     0.		0.001     -0.05   0.05
par   c3		  nui     0.		0.001     -0.05   0.05
par   c4		  nui     0.004		0.001     -0.05   0.05
par   c5		  nui     0.004		0.001     -0.05   0.05
par   Aps100x100          nui     2.5E-04	1E-05     0.0     0.1
par   Aps100x143          nui     1.1E-04	7E-06     0.0     0.1
par   Aps100x217          nui     9.9E-05	6E-06     0.0     0.1
par   Aps143x143          nui     4.7E-05	2E-06     0.0     0.1
par   Aps143x217          nui     3.1E-05	3E-06     0.0     0.1
par   Aps217x217          nui     7.6E-05	6E-06     0.0     0.1
par   Asz                 nui     1		0.1       0.0     10
par   Acib                nui     1.		0.1       0.0     10
par   AdustTT		  nui	  1	         0.1	  0.0     2
fix   AdustPP		  nui	  0.00         0.1	  0.0     2
fix   AdustTP		  nui	  0.00         0.1	  0.0     2
par   Aksz                nui     0.00         1.0	  0.0     10
par   Aszxcib             nui     0.00         1.0	  0.0     10
##########################################################################
\end{lstlisting}    

Next you may add some external constraints (or "priors" in Bayesian vocabulary). The most common one is the 1D Gaussian, which is illustrated here: for instance we add a Gaussain likelihood to "AdutTT" centered on 0 with $\sigma=0.2$.

\begin{lstlisting}[caption=Adding Gaussian priors]
#priors######################################
gauss1    AdustTT  1 0.2
gauss1    Acib     1 0.2
gauss1    A_planck 1 0.0025

gauss1    c0        0     2e-3
gauss1    c1        0       2e-3
gauss1    c3        0       2e-3
gauss1    c4        0.0025      2e-3
gauss1    c5        0.0025      2e-3
###############################################
\end{lstlisting}

Then you may need to tune the details of the Boltzman solver. For \class you can do it with the keyword {\tt class} which passes the following argument as a strings to the solver. For instance to define the non-cold-dark matter setup to a degenerate massive neutrino with 0.06\eV\ (the upper limit in normal hierarchy measured from oscillation experiments) one may use:
\begin{lstlisting}[caption=Defining extra \class setup]
#class setup###############################
class     N_ncdm                1
class     m_ncdm                0.06
class     N_eff                 2.046
###########################################
\end{lstlisting}
Note that in \class, $\neff$ corresponds only to the pure radiation today which is why it is not the famous 3.046 value). 
One may also pass these parameters as previously with the "fix" keyword, but this better decouples the variables from their environment (and avoids repeating fixed numbers in the output files). There are some cases where strings to \class 
is mandatory:
\begin{lstlisting}[caption=Direct transmission to \class of string parameters]
##################################################
class     k_pivot               0.05
class     lensing               yes
class     sBBN\ file            bbn/sBBN.dat
#################################################
\end{lstlisting}
(Note the \verb|\| in the last line to indicate a trailing space)

Finally we define the file that contains some high-precision settings when we wish to compute precisely the MLE or profile with \class. Indeed as we will see later getting a precise $\chi^2$ minimum is very challenging with a method that cannot rely on analytical gradients.
More exactly Boltzmann solvers do not lead to a likelihood that is continuous in the parameters space because they are the result of many numerical computations with some limited accuracy (as cutoffs in the integrals).~\class, beside being written in clean \text{C}, is interesting in this respect since it allows to specify in a single place all the precision parameters that are used. We provide a set of pre-defined precision files (which result from a trade-off between accuracy and CPU-time and are presented in \citet{prof14}) and suggest to use the following one:

\begin{lstlisting}[caption={Using a higher precision setup for \CLASS\ computations},label=hpjul2]
#########################################
precisionFile=class_pre/hpjul2.pre
#########################################
\end{lstlisting}
when performing MLE or profile-likelihood studies. Default settings can be used when doing MCMC, because the algorithm smooths the details of the likelihood function. More in the next section.

The parameter file is ready and we call it \parfile. 
To test it you can run 
\begin{verbatim}
writeChi2 hlpTT_bflike_LCDM.par
\end{verbatim}
which parses your file and just writes the $\chi^2$ value corresponding the the given parameter value (no fit there).
If the program goes to the end, you most probably have some valid file.

\subsubsection{Adapting the Boltzmann solver precision}
\label{sec:precision}
\begin{figure}[h!]
\begin{center}
\includegraphics[width=0.7\columnwidth]{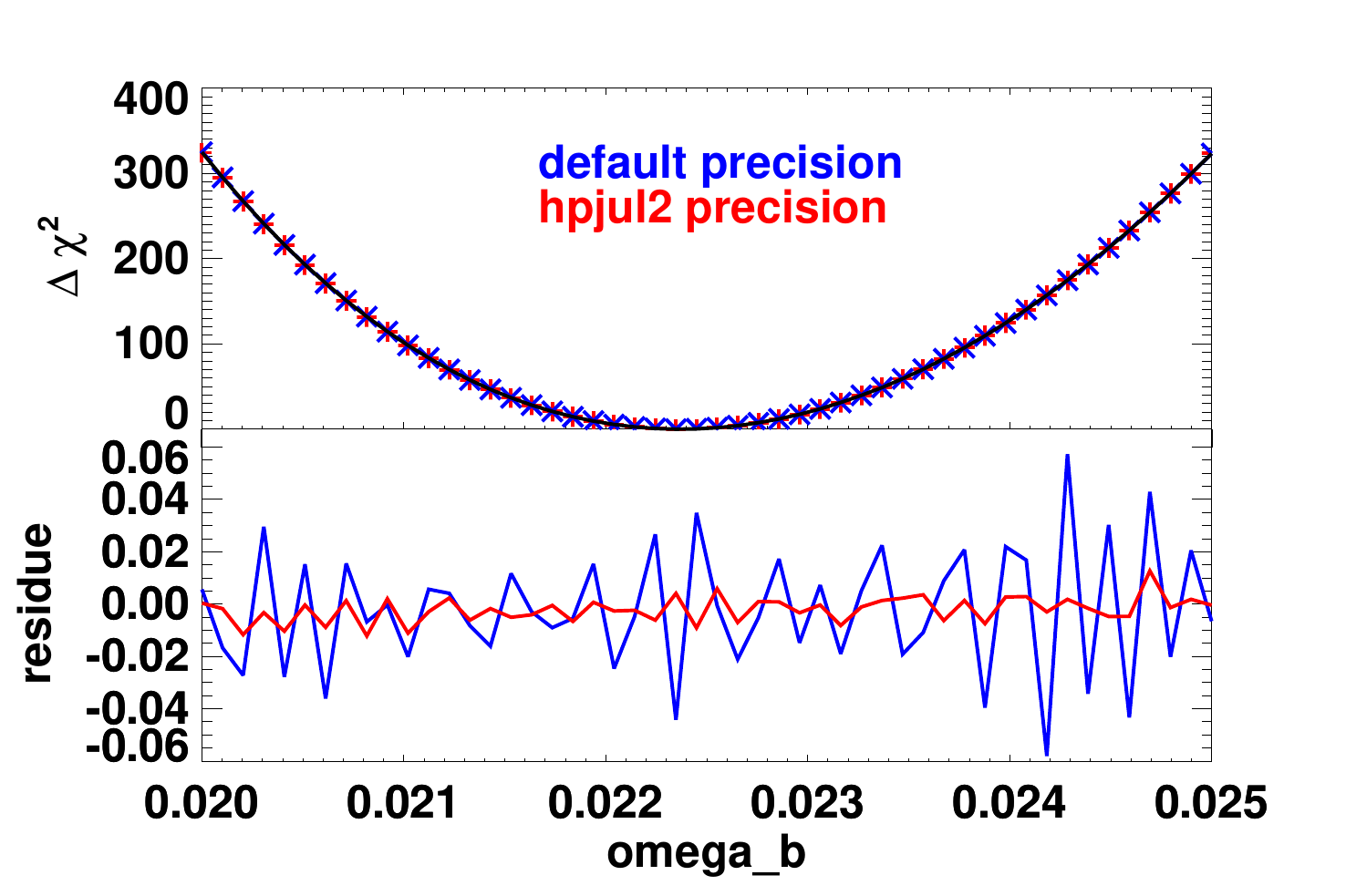}
\caption{{\label{fig:scanomb}
The upper plot shows the $\chi^2$ values obtained from the {\tt ScanParam} program using the \parfile setup, in blue when not specifying any precision file for \class and in red using the hpjul2 one. A smooth fit is superimposed in black. The lower plot shows the residuals in both configurations and reflects the numerical noise of the Boltzmann Engine.%
}}
\end{center}
\end{figure}

While in practice using the \verb|hpjul2.pre| is generally sufficient, we show here
how you can get an idea of the numerical noise in your analysis.
We recall that the difficulty for finding a really precise multi-dimensional $\chi^2$ minimum comes from the fact that the function $\dfrac{\partial\lik}{\partial\Omega_i}$ is not continuous because we use a full Boltzman solver to compute the observable $X$ in $\lik(X(\Omega))$.
We can get some idea of the level of numerical noise, by scanning linearly one variable (with all other ones fixed) and studying the smoothness of the $\chi^2(\theta_i)$ values.
This is the goal of the \verb|ScanParam| program which fixes all but one parameters from the file,  vary linearly the free one and records the $\chi^2$ values in a text file.
Let's see how this works for instance on the parameter $\omb$. Run:
\begin{verbatim}
ScanParam hlpTT_bflike_LCDM.par omega_b 0.020 0.025 50
\end{verbatim}
where the last 3 values indicates how the variable is scanned (here from 0.02 to 0.025 with 50 equidistant points).
The upper plot of Fig. \ref{fig:scanomb} shows the $\chi^2$ values, in blue when using the default \class precision parameters (ie. no particular precision file) and in red with the \verb|hpjul2.pre| one.
Then the black curve shows a smooth interpolation between the points, and the bottom plot the residual of each point w.r.t to it. We then see that the numerical noise with the \verb|hpjul2.pre| file (in red) is much lower than for the default case (in blue). Using such a precision allowed in all cases we studied to determine an accurate $\chi^2$ minimum.

\subsubsection{Estimating the wall time}
\label{sec:time}
It can be useful before running jobs to estimate the time taken by each iteration. By iteration we mean the complete processing of one model. For this you can run interactively a few iterations of the \verb|Minimize| executable adding some verbosity to your parameter file with: 
\begin{lstlisting}
verbose=true
\end{lstlisting}
This is done with 
\begin{verbatim}
Minimize myparfile output
\end{verbatim}
where "output" is any name.

It will print among many other information the time used for each iteration (at the end of each line).
The CPU time for \parfile typically varied between 5s on 8 cores to 2s on 24 ones  (with the \verb|hpjul2| precision file) and is only limited by the scalability of the Boltzman solver.

\subsection{\label{sec:MLE} The Maximum Likelihood Estimator (MLE), aka "best fit"}
The MLE enjoys many nice properties that can be reviewed e.g in \citet{jamesbook}.
One that is not always well known (and that also applies to the likelihood ratio statistics) is the MLE invariance which states that if $\hat \theta$ is the MLE of $\theta$ , then the MLE of any function of $\theta$, $y=f(\theta)$ is $\hat y=f(\hat \theta)$. 

For instance in the \lcdm\ model, there are 6 free cosmological parameters often chosen as 
$\Omega={\omb, \omc, \theta, \ns\, \lnAs, \tau\\}$ (see \parfile). If instead of the angular size of the sound horizon ($\theta$) one uses the Hubble constant ($H_0$), the minimization leads exactly to the same values for all the other parameters. Furthermore, if the minimization was performed on the $\Omega$ set, one does not need to redo it with $H_0$. From the invariance property you just need to run the Boltzmann solver on the $\hat \Omega$ solution to obtain the $\hat H_0$ best fit solution. 

\subsubsection{Minimisation}
\label{sec:mini}
The goal is to measure in the parameter space where the likelihood reaches its maximum value or equivalently the $\chi^2$ its minimum (which is preferred for numerical reasons).
As discussed in Sect.\ref{sec:precision} obtaining a precise multi-dimensional minimum (here on 23 parameters but it can be more) is challenging since
\begin{itemize}
\item we don't have analytical first order derivatives,
\item Boltzmann engine computations brings up numerical noise.
\end{itemize}

Many softwares exists for multi-dimensional optimization.
We chose to  work with the \minuit \footnote{\wwwminuit} package, a renowned
minimizer often used in High Energy Physics. Starting in the 70's \citep{minuit}, it was 
permanently improved and rewritten into \texttt{C++}. This version, named \texttt{MINUIT2} is embedded within \CAMEL.
\minuit is a large toolbox, including many features (as releasing or constraining parameters on the fly), and working with it requires some degree of experimentation. We tuned it for cosmology with the {\tt MIGRAD} algorithm that is a "variable metric method" (one changes the metric of the system not the data, see eg. \citet{vmm}).
In most cases you won't need to change the default setting.

To run the minimization as a batch job, one uses the \verb|Minimise.sh| script (note the ".sh"):
\footnote{Scripts are located under the \texttt{work/batch} directory and we propose 
some implementations for cc-inp3 (\url{cc.in2p3.fr}) and NERSC (\url{nersc.gov}); they should be simple to adapt to other sites.}

\begin{verbatim}
Minimize.sh /path/to/hlpTT_bflike_LCDM.par
\end{verbatim}
This
\begin{enumerate}
\item checks your configuration,
\item creates a directory named hlpTT\_bflike\_LCDM\_min where you launched the script, which will contain the outputs,
\item sends the (OpenMP) job.
\end{enumerate}

You then wait for the job completion...
In our case, the minimization converged in about 3000 iterations, which, given the wall times measured in Sect.\ref{sec:time} correspond typically to 1.5h (on 24 cores) or 4h on 8. This time is only limited by \class\ computations not the minimization algorithm. 

One then obtain the job log-file and the {\tt bestfit.dat} and {\tt covmat.data} files. You can examine the log-file and not worry too much about \class\ error messages (if they are sparse!). They are often due to the lack of convergence of some numerical algorithms for some "extreme" combination of parameters: \camel\ catches these errors and the minimizer moves away from those regions.
You should also not worry about such a message:
\begin{verbatim}
WARNING: Minuit did not converge.
\end{verbatim}
which is due, for \class, to the numerical discontinuities discussed in Sect \ref{sec:precision}.
Then near the end you will find the interesting results.
\begin{verbatim}
# ext. ||   Name    ||   type  ||     Value     ||  Error +/-

0   ||   omega_b   ||  limited ||0.02221669365664     ||0.0001757676645765
1   ||   omega_cdm ||  limited ||0.1193778692005      ||0.001014938068844
2   ||   100*theta_s||  limited ||1.041727722902       ||0.000349347668509
3   ||   tau_reio  ||  limited ||0.07124770721867     ||0.002294613708248
4   ||   log(10^10A_s)||  limited ||3.068270305092       ||0.00899043822157
5   ||   n_s       ||  limited ||0.9657674074055      ||0.006549911640786
6   ||   A_planck  ||  limited ||1.000680632922       ||0.002480642264396
7   ||   c0        ||  limited ||0.001487875696301    ||0.001638396021085
8   ||   c1        ||  limited ||0.001440151762443    ||0.00163998060017
9   ||   c2        ||   const  ||0                    ||
10  ||   c3        ||  limited ||-0.0007967689027367  ||0.00147758139892
11  ||   c4        ||  limited ||0.001778114967395    ||0.001762250083468
12  ||   c5        ||  limited ||0.001921014220322    ||0.001759624112839
13  ||   Aps100x100||  limited ||0.000255535488113    ||3.31867430043e-05
14  ||   Aps100x143||  limited ||0.0001108863113654   ||2.529138206055e-05
15  ||   Aps100x217||  limited ||7.71374529739e-05    ||1.641760577019e-05
16  ||   Aps143x143||  limited ||4.429855974166e-05   ||2.030311938827e-05
17  ||   Aps143x217||  limited ||3.222589327638e-05   ||1.452202004259e-05
18  ||   Aps217x217||  limited ||7.645366758567e-05   ||1.374833402066e-05
19  ||   Asz       ||  limited ||1.322514028996       ||0.5973930819106
20  ||   Acib      ||  limited ||0.9397363536028      ||0.1319397001748
21  ||   AdustTT   ||  limited ||0.9832158679645      ||0.09576457283827
22  ||   AdustPP   ||   const  ||0                    ||
23  ||   AdustTP   ||   const  ||0                    ||
24  ||   Aksz      ||  limited ||1.634684530752       ||5.729577959257
25  ||   Aszxcib   ||  limited ||8.295195641495e-08   ||8.899445674664
\end{verbatim}
which gives the values of the best fit and some (crude) estimate of the "error bars" from the Hessian matrix (that should not be trusted too much, more in Sect. \ref{sec:hessian}).
Also of uttermost importance is the value of the $\chimin$ at this location which appears as 
\begin{verbatim}
minimum function Value: 20448.12723918
\end{verbatim}
and at the very end of the file you obtain the breakdown of the individual \chimin values and the total wall-time. Here we obtain:
\begin{verbatim}
===== lowl_SMW_70_dx11d_2014_10_03_v5c_Ap.clik: chi2=10495
===== HiLLiPOP:DX11dHM_superExt_CO_TT.lal: chi2=9951.47
Chi2 from HiLLiPOP: 9951.47 ndof: 9556
===== CMB_all: chi2=20446.5
===== Gauss1 constraint on AdustTT(1+-0.2): chi2=0.00704268
===== Gauss1 constraint on Acib(1+-0.2): chi2=0.0907927
===== Gauss1 constraint on A_planck(1+-0.0025): chi2=0.0741218
===== Gauss1 constraint on c0(0+-0.002): chi2=0.553444
===== Gauss1 constraint on c1(0+-0.002): chi2=0.518509
===== Gauss1 constraint on c3(0+-0.002): chi2=0.15871
===== Gauss1 constraint on c4(0.0025+-0.002): chi2=0.13028
===== Gauss1 constraint on c5(0.0025+-0.002): chi2=0.0838061
TIMER TOTAL TIME=16268.6 s      =271.144 min    =4.51907h
===== Chi2Combiner: chi2=20448.1
\end{verbatim}

Now for simplicity we included the {\tt bestfit.awk} script \footnote{analysis scripts are located in the \texttt{work/tools} directory} which parses the whole output file and prints out the interesting things.
Use it as:
\begin{verbatim}
awk -f tools/awk/bestfit.awk logfile
\end{verbatim}
The files {\tt bestfit.dat} and {\tt covmat.dat} are also interesting excerpts of the log-file that can be easily read with any interactive analysis software.

\subsubsection{Trying different starting points}

It can be wise to test whether one could not get a slightly better (lower) $\chimin$ value through a different path (recall the numerical Engine noise may perturb the gradient descent). To this end, we would like to shoot different starting points and see what we obtain. This can be done specifying a job "range" in the form "I-J". For example:
\begin{verbatim}
Minimize.sh path/to/hlpTT_bflike_LCDM.par 1-10
\end{verbatim}
creates the {\tt hlpTT\_bflike\_LCDM\_bf/} directory (that contains everything) and sends 10 jobs to the cluster through what is referred to as an "array-job".

Each job is independent, so can be run simultaneously with all the others finally increasing little the user wall-time.
Job number 1 is the standard minimization described previously. Others will perform some uniform random shuffle of each parameter within the [min,max] bounds of  the parameter file.
These shots are reproducible in the sense that another "1-10" run would yield the same configuration. Use "11-20" if you want 10 more trials...

Now we have the outputs of jobs 1 to 10 in the same directory. Which is the best? the one with the lowest \chimin.
We provide a script in order to help finding it. Go into  {\tt hlpTT\_bflike\_LCDM\_bf/}
and run
\begin{verbatim}
 awk -f tools/awk/zebest.awk *.o*
 \end{verbatim}
where "*.o*" is a pattern of all the log-files files (and may be different at your site). Here is an example of what you can obtain (given the random seed do not expect exactly the same)
\begin{verbatim}
hlpTT_bflike_LCDM_bf.o26551546.1 chi2=20448.127239 it=2734 t=5.950475s tot=4.519056h  errors=2
hlpTT_bflike_LCDM_bf.o26551546.10 chi2=20448.205740 it=4706 t=6.512537s tot=8.513333h  errors=13
hlpTT_bflike_LCDM_bf.o26551546.2 chi2=20448.276149 it=3191 t=5.881981s tot=5.213722h  errors=15
hlpTT_bflike_LCDM_bf.o26551546.3 chi2=20448.148907 it=5165 t=6.357231s tot=9.120861h  errors=20
hlpTT_bflike_LCDM_bf.o26551546.4 chi2=20448.199447 it=3300 t=6.380727s tot=5.849000h  errors=56
hlpTT_bflike_LCDM_bf.o26551546.5 chi2=21113.587792 it=1721 t=5.421063s tot=2.591569h  errors=56
hlpTT_bflike_LCDM_bf.o26551546.6 chi2=20448.158418 it=3819 t=5.074077s tot=5.382750h  errors=57
hlpTT_bflike_LCDM_bf.o26551546.7 chi2=20448.667074 it=2867 t=6.624834s tot=5.275944h  errors=86
hlpTT_bflike_LCDM_bf.o26551546.8 chi2=20448.154056 it=3971 t=5.468623s tot=6.032194h  errors=86
hlpTT_bflike_LCDM_bf.o26551546.9 chi2=20448.225423 it=2925 t=3.550632s tot=2.884889h  errors=87
--------------------------------------------------------
chi2best= 20448.12723918  for hlpTT_bflike_LCDM_bf.o26551546.1
\end{verbatim}
For each line you find : its name, \chimin, number of iterations, time per iteration, total time, number of (possible) \class errors (warnings).
So in our case, the best result was indeed for the first job, ie. the one analyzed in Sect. \ref{sec:mini}.
Some "accidents" may happen (large $\chi^2$ as for the 5th job) where some extreme random starting values did not allow a proper convergence.
But basically we are confident the best fit initially found is indeed the best one.

\subsubsection{The Hessian matrix}
\label{sec:hessian}
\minuit allows to retrieve the Hessian matrix corresponding to the local curvature of the parameters around the minimum.
It is produced when adding a name for the matrix to the arguments of the \verb|Minimize| executable (which is the default in the \verb|Minimise.sh| script)
You should be careful when using it since:
\begin{itemize}
\item it is not very precise due to the numerical noise,
\item it corresponds to second-order partial derivative matrix and cannot be identified to an "error matrix" in the case of strongly non linear parameters \citep{jameserrors}.
\end{itemize}
It should therefore only be considered as indicative: the right way to obtain real "errors" is through the use of the profile-likelihood or MCMC methods.
As we will see in Sect \ref{sec:MCMC}, it is however very convenient as the input proposal to the adaptive MCMC algorithm and we will check {\it a posteriori} that it was quite precise in this case (Fig. \ref{fig:cors}).

\subsection{Profile-likelihoods}
\label{sec:profile}
\subsubsection{Properties}
We now address the question of\textit{ interval estimation} for a single parameter (or a pair, rarely more) using the method of profile-likelihoods. Technically it consists in \textbf{fixing} values of a given parameter (let's call it $\theta_i$) and performing a minimizations w.r.t all the others. For each $\theta_i$ scanned value, the $\chi^2(\theta_i)=\chimin$ value is reported on a graph that is finally offset to 0. We will focus our discussion on 1D profile-likelihoods, ie. when one scans a single parameter, but 2D ones (i.e scanning values over a grid) are also available in \camel.

The interest of reconstructing a profile-likelihood stems from the fact that by cutting it, one can reconstruct an $\alpha$-level confidence intervals in the frequentist sense: if the experiment is repeated many times the interval covers the true values a fraction $\alpha$ of the times. 
We first consider the case of a Gaussian variable. 
Then $\Delta \chi^2$, considered as random variable of the data, is centered on the true value and follows a  $\chi^2_1$ distribution  \footnote{or $\chi^2_2 $ for 2 parameters} (see eg. \citet{Severini}). Then the $\alpha$-level interval is obtained by:
\begin{align}
  P(\Delta \chi^2) & < \chi_{\alpha} \\
 \text{where}~\int_0^{\chi_\alpha} & \chi^2_1(t) dt = \alpha .
\end{align}

We give in Table \ref{tab:cutprof} some classical values  of $\chi_\alpha$ for 1 and 2D profile-likelihoods.

\begin{table} 
\centering
    \begin{tabular}{ c|c|c|c }
         dim  & 0.68 & 0.95 & 0.99 \\
         \hline
         1 par & 1 & 3.84 & 6.63\\
         2 par & 2.29 & 5.99 & 9.21 \\
         \hline
    \end{tabular} 
   \caption{{\label{tab:cutprof} Cutoff values to apply to $\Delta\chi^2$ for 1 or 2 parameters ("dim")  in order to obtain  68,95 and 99\% CL intervals}}
\end{table}

Although we built our discussion from the Gaussian case, it \textit{actually holds for non-Gaussian cases too} (ie. non parabolic $\Delta \chi^2$ curves) and allows to reconstruct asymmetric intervals (w.r.t the MLE).
As discussed in \citet{jamesbook}, in the general case one can always \textit{imagine} a change of variable that would make the profile parabolic. The MLE invariance discussed Sect \ref{sec:MLE} more generally applies to likelihood ratios $\dfrac{\lik(\theta_i)}{\lik_{max}}$ or equivalently $\chi^2(\theta_i)-\chimin$. Thus the assertion made on the interval for the transformed variable ($\Delta \chi^2=1$) also applies to the original variable. Strictly speaking there can be a slight difference (because one transforms the data not the true value) but this only happens for very low statistics (a data samples) which is anyway always difficult to treat and a situation where the MLE is not necessarily optimal.

For completeness, we note that there are some cases where one wants to incorporate some physical bound (as for example a mass being always positive). This is most naturally performed in the Bayesian framework through the use of priors. The "unified approach" of \citet{FC} can be applied here and was demonstrated in \cite{prof14} and \cite{Henrot_Versill__2015}.

Before showing a concrete example we would like to point out that because of the method, the best fit at the minimum of the $\Delta \chi^2(\theta_i)$ function is \textit{the best-fit of the full set of parameter}s (the one obtained in Sect.\ref{sec:MLE}).
This provides an interesting consistency test and ensures that, unlike in the MCMC approach, there is no difference when quoting the parameter value and the best fit solution.

\subsubsection{Profile-likelihoods in practice}

Building a profile-likelihood is then nothing more than running several minimizations fixing one parameter each time.
Although there exist a \camel executable that loops on the variable and perform sequentially minimizations, one can take advantage of batch farms by sending each minimization on a separate worker.
This can be performed with the \verb|Profile.sh| script.
To use it, one must first write a \textit{scan-file} which specifies the profiled variable and its values. For instance, in a file \verb=omega_b.scan= we put \footnote{this is only illustrative, you rarely need so many points. Note also that since this file is sourced by the script, one can use some \texttt{bash} commands (as loops) to generate these numbers.}
\begin{verbatim}
var="omega_b"
val=(0.022000 0.022020 0.022041 0.022061 0.022082 0.022102 0.022122 0.022143 0.022163 0.022184 
0.022204 0.022224 0.022245 0.022265 0.022286 0.022306 0.022327 0.022347 0.022367 0.022388 0.022408 
0.022429 0.022449 0.022469 0.022490 0.022510 0.022531 0.022551 0.022571 0.022592 0.022612 0.022633 
0.022653 0.022673 0.022694 0.022714 0.022735 0.022755 0.022776 0.022796 0.022816 0.022837 0.022857 
0.022878 0.022898 0.022918 0.022939 0.022959 0.022980 0.023000)
\end{verbatim}

\begin{figure}[h!]
\begin{center}
\includegraphics[width=0.7\columnwidth]{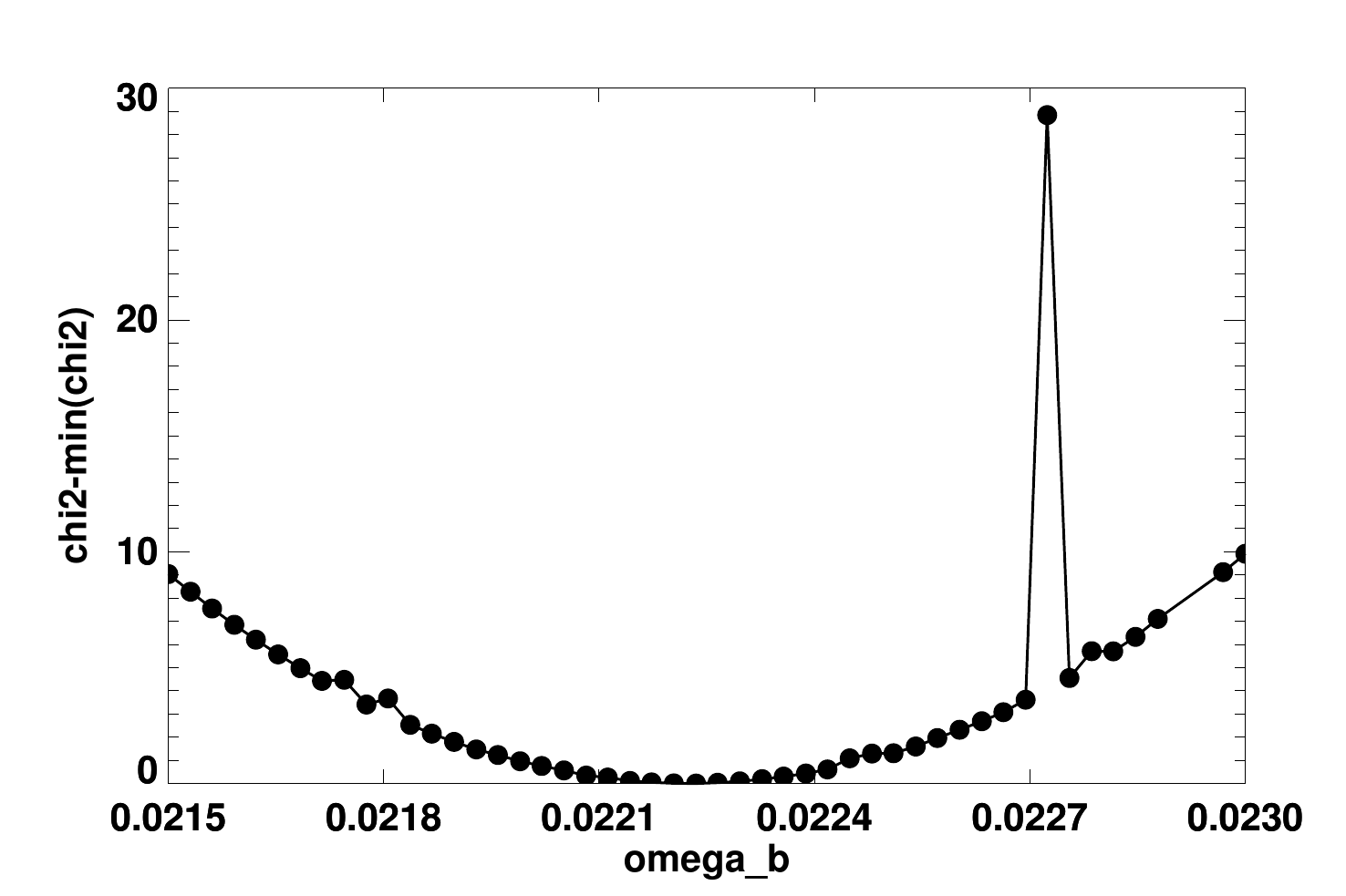}
\caption{{\label{fig:prof_omegab1}
A first reconstruction of the omega\_b profile likelihood points using \parfile with one single starting point%
}}
\end{center}
\end{figure}

Then run:
\begin{verbatim}
Profile.sh path/to/hlpTT_bflike_LCDM.par /path/to/omega_b.scan
\end{verbatim}
When all the jobs finished, we concatenate the best-fit files \footnote{we will see later how to do that.} and represent the "chi2min" vs. "omega\_b" variables reading the corresponding columns of the concatenated file (the profiled variable is always the column before the "chi2min" one). We obtained the points shown on Fig. \ref{fig:prof_omegab1}.

Obviously there has been a "glitch" near 0.0227 and some "wiggles" indicates the fit did not reached exactly its minimum in 3 or 4 places. As for the best fit case (Sect \ref{sec:mini}) to cure that issue we shoot different starting points. This does not need to be re-performed for each point, only the ones that are manifestly incorrect: this is the reason for writing explicitly the scan-file: you may re-edit it and keep only the values that needs to be improved.

Here we re-run (for the sake of illustration) on all the point (ie keep the same \textit{scanfile}) shooting each time 6 different starting points for minimization:
\begin{verbatim}
Profile.sh path/to/hlpTT_bflike_LCDM.par /path/to/omega_b.scan 1-6
\end{verbatim}
"1-6" means that jobs 1 to 6 will be launched for each scanned value. Each one (except for number 1) will shuffle randomly the starting values of each parameter in their specified [min,max] range. This shot is reproducible, so that if you need more starting points do not use "1-6" again but something as "7-10". This allows to increase gradually the statistics until reaching satisfactory continuity.

\begin{figure}[h!]
\begin{center}
\includegraphics[width=0.7\columnwidth]{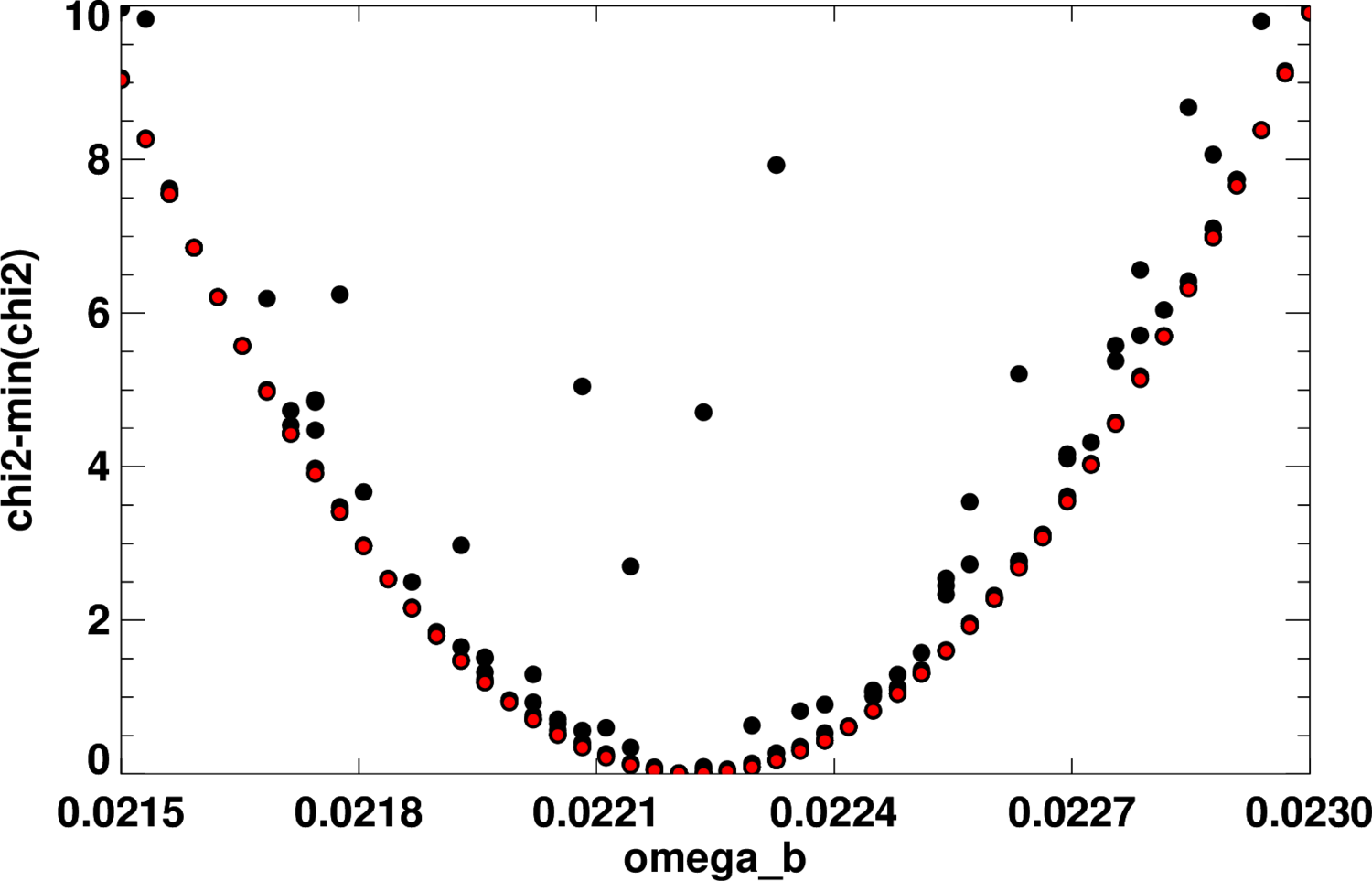}
\caption{{\label{fig:mprof_omegab}
\dchi minima reached with different starting points (6) for $\omb$ scanned values. The profile likelihood is obtained as the minimal values for each abcissa point and shown in red.%
}}
\end{center}
\end{figure}

The script creates the \verb|hlpTT_bflike_LCDM_mprof_omega_b/| directory and within it several other ones corresponding to the scanned values (which will contain each the outputs of the 6 jobs).
Let us look at the outputs of all the jobs by concatenating all the \verb|best_fit| files.
All the points are shown on Fig. \ref{fig:mprof_omegab}.

Some fits did not converge well, but what we are interested in is, for each scanned value, the minimal $\chi^2$ value.
There are in each case our "best best fit" solution and shown in red on Fig. \ref{fig:mprof_omegab}.
Using them the profile is now nicely continuous.
For convenience, and since it is a common operation, we provide a script (in \verb|work/tools/python|) to extract these points. Simply run it on the list of sub-directories containing the scanned values. Here we used (in the directory 
\verb|hlpTT_bflike_LCDM_mprof_omega_b|)
\begin{verbatim}
python tools/python/bestbestfit.py omega_b*
\end{verbatim}
where \verb|omega_b*| is the pattern of the output directories. This creates the \verb|profile_omega_b.txt| file.

\begin{figure}[h!]
\begin{center}
\includegraphics[width=0.7\columnwidth]{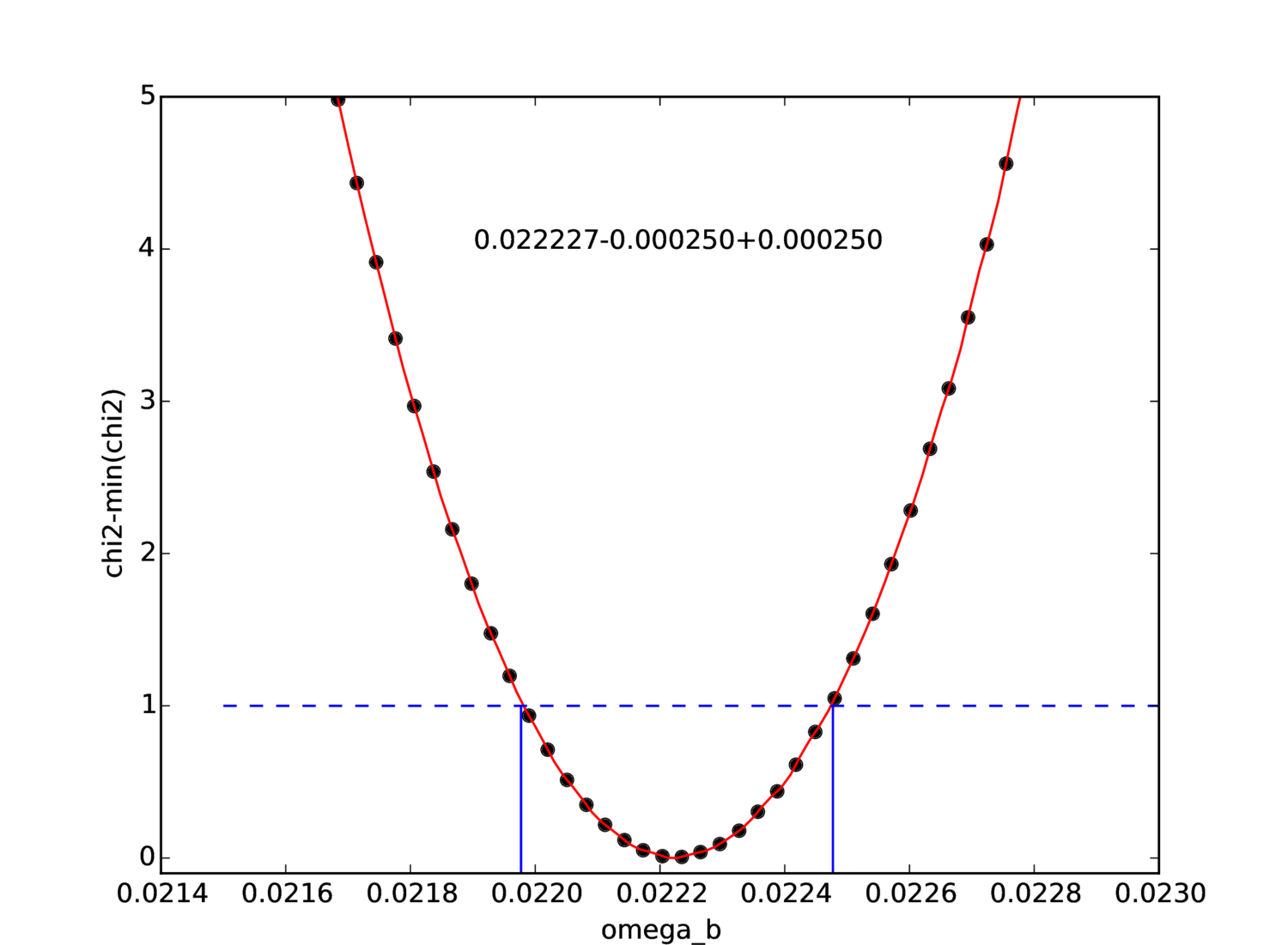}
\caption{{\label{fig:bestprof_omegab}
Construction of the 68\%CL interval by thresholding the interpolated $\omb$ profile-likelihood (in red) at 1.%
}}
\end{center}
\end{figure}

Then you interpolate smoothly between the profile points and determine the intersection with the $\dchi=1$ line to obtain the 68\% CL interval. You may quote "error bars" by comparing these limits to the position of the minimum. Be careful however to determine the minimum and limits on the interpolated curve not the data points.
You can code this yourself in your preferred environment or use the \verb|draw_profile.py| script which produced Fig. \ref{fig:bestprof_omegab} running:
\begin{verbatim}
python tools/python/draw_profile.py profile_omega_b.txt 
\end{verbatim}

There are other ways to produce this plot (and others) through the \verb|tools/python/camel.py| library (see the \camel\ web site for details).

\begin{figure}[h!]
\begin{center}
\includegraphics[width=1\columnwidth]{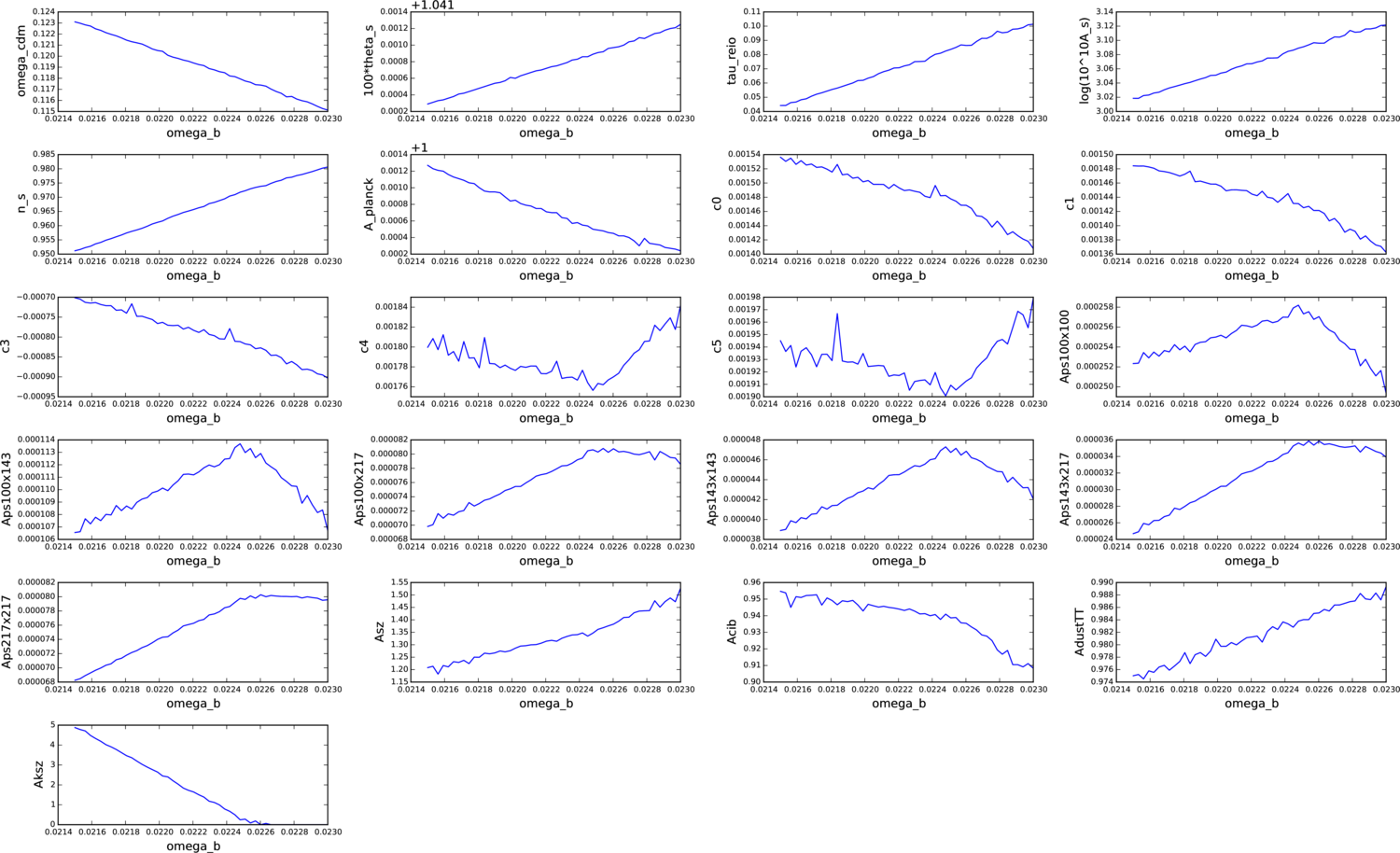}
\caption{{\label{fig:coprof_omegab}
Each plot shows the value of the $\theta_i$ parameter w.r.t to the scanned variable ($\omb$) at the best-fit solutions. We call these curves "co-profiles". They are an indication whether the 2D profile-likelihoods are Gaussian (in which case they should be slopes).%
}}
\end{center}
\end{figure}

There is actually more information we can extract from the set of best-fits. For each fixed $\omb$ value we have the multi-dimensional solution of the minimum. Then one can look not only at the $(\omb,\dchi)$ curve 
but also at the best-fit solution achieved for the other variables ie. $(\omb,\theta_i)$. This represents the covariance between $\omb$ and all the other variables ( but not the \textit{correlation} since we are missing $\sigma_{\theta_i}$). These curves represents the $(\omb,\theta_i)$ minimum valley of the likelihood and are similar (but not equivalent since there is no marginalization) to the major-axes of the 2D Gaussian posteriors in an sampling approach. In any case inspecting these "co-profile" is a good way to check that none of the parameters was blocked at some limit.
We provide the python scritpt \verb|coprofiles.py| to produce these plots.
We ran:
\begin{verbatim}
python tools/python/coprofiles.py profile_omega_b.txt
\end{verbatim}
and it produced the output shown on Fig. \ref{fig:coprof_omegab}.

\subsubsection{Profile-likelihood summary}
The construction of a profile-likelihood may seem delicate because we entered in many details. In practice you should pay attention to the following points:
\begin{itemize}
\item always use a precision file (with \class). the default \verb|hpjul2.pre| should be sufficient in most cases.
\item choose your scanning points carefully. You need to focus only on the $\dchi \lesssim 5$ region (which may require some iterations): it is useless (and difficult) to regularize points outside.
\item although we described the method with many points (50), you don't need as much to construct a reliable curve (10 to 20 are sufficient).
\item you should first try with the parameter-file fixed starting point (ie. range 1-1 in \verb=Profile.sh= which is the default)
\item then if you notice some continuity problem
\begin{itemize}
\item if there are little, simply discards those points before performing a smooth interpolation among the remaining "good" ones.
\item if there are many, try more starting points on the problematic values
\end{itemize}
\item once you are happy with your profile, always check the "co-profile" to see if some parameters are not blocked by the [min,max] bounds.
\end{itemize}

\subsubsection{Results on \lcdm\ cosmological parameter}
\begin{figure}[h!]
\begin{center}
\includegraphics[width=1\columnwidth]{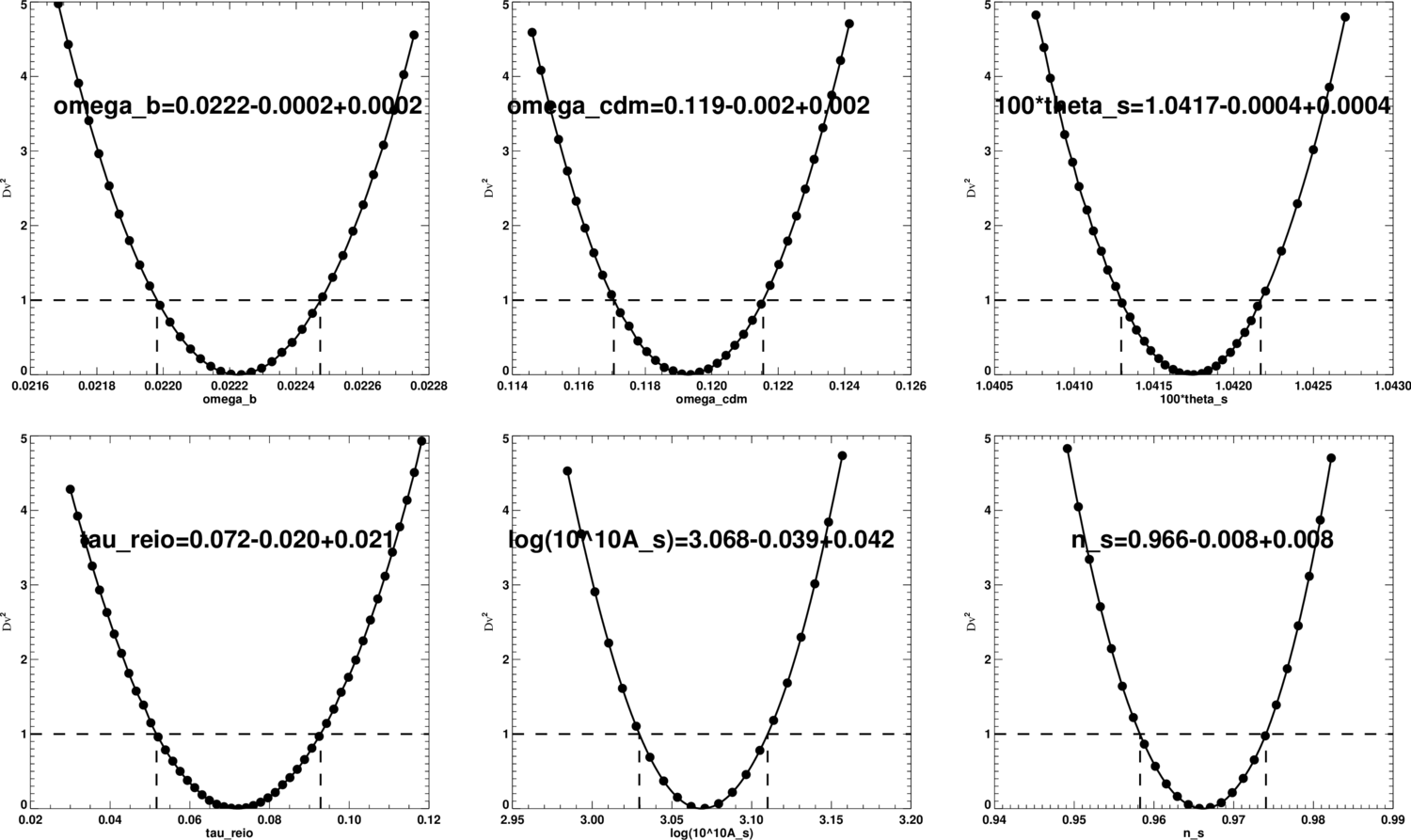}
\caption{{\label{fig:lcdm_prof}
Profile-likelihood reconstructed for the 6 \lcdm\ cosmological parameters with the \parfile\ parameter file. The line at 1 gives the 68\% CL region of each parameter.%
}}
\end{center}
\end{figure}

We apply this methodology to the remaining 5 cosmological parameters in the file \parfile and obtain the profile-likelihoods shown on Fig. \ref{fig:lcdm_prof}

\subsection{\label{sec:MCMC} MCMC}
\subsubsection{Real-life Metropolis}

The Monte-Carlo Markov Chain method draws samples according to any distribution (for a nice introduction see eg. \citet{Bardenet_2013}).
When applied to a likelihood function, it may serve to reconstruct the unknown parameters distribution (posteriors), a concept not accepted in a classical "frequentist" perspective.
According to Bayes theorem:
\begin{align}
p(\theta|x)=\dfrac{p(x|\theta)\pi(x)}{p(x)} \propto \lik(\theta|x)\pi(\theta)
\end{align}
$\lik$ being the likelihood of the data and $\pi$ the prior distribution that reflects the degree of belief on the parameter before considering the experiment. The latter is rarely discussed (ie. considered as equal to 1 in some reasonable range although this is not necessarily the most non-informative prior \citep{Jeffreys}) in which case the likelihood of the parameters equals their probability. Sampling from the likelihood then gives directly the parameters posterior distribution.

The founding father of all MCMC algorithms is the Metropolis one \citep{metropolis49,metropolis53}. Its mathematical properties have been studied for decades and it is certainly the best characterized one \citep{Roberts,gilks}.
Although it is {\it mathematically} guaranteed to converge to the proper distribution at infinity, this is often too far in real case studies, in particular in cosmology, where Boltzmann solver computations are slow.
In practice it means that the {\it proposal distribution} used to perform the random jumps, generally a central multi-variate Gaussian with some covariance matrix,  must be known precisely. This matrix is often complicated because of important off-diagonal elements that can hardly be a priori inferred (see eg. Fig. \ref{fig:cors}): not incorporating them will not allow a proper sampling of the likelihood on a reasonable timescale. Therefore the matrix is reconstructed iteratively from the empirical covariance of the samples but it is a very time-consuming process. \textit{ Reproducing} results when the proposal is \textit{given} is simple but hides the lengthy previous steps. A soon as one want to study new variables/likelihoods the proposal matrix must be rebuild again.

We have therefore implemented in \camel an adaptive scheme of the Metropolis algorithm. In this approach the proposal covariance matrix is built on the fly: the adaptation level is gradually decreased so that one ends up with a fixed matrix and then follows the classical Metropolis algorithm. The initial matrix does not need to be known precisely and can be obtained in 2 ways:
\begin{enumerate}
\item one may use the Hessian matrix obtained from minimization (Sect \ref{sec:hessian}).
\item often one just want to expand a set of parameters, and already knows from the empirical covariance a precise matrix of a subset of parameters: then just add a crude estimate of the new parameters errors on the diagonal and let the algorithm do the work.
\end{enumerate}

\subsubsection{Adaptive Metropolis (AM)}
The original algorithm \citep{Haario_2001} works in the following way:

\begin{enumerate}
\item Choose a starting point $X_0$,  a starting covariance matrix $\Sigma_0$ and a tuning parameter $c$;
 \item generate a value $Y$ from a proposal density $N(X_{t-1},c\Sigma)$;

\item evaluate the test ratio $\alpha=min[1,\frac{\pi(Y)}{\pi(X)}]$;

\item generate a value $u$ uniformly distributed in $[0,1]$;

\item if $u \leq \alpha(X_t,Y)$ set $X_{t+1}=Y$ else set $X_{t+1}=X_{t}$;

\item  update running mean and covariance:
\begin{equation}
 \mu_t=\mu_{t-1} + \frac{1}{t}(X_t-\mu_{t-1}),~~~\Sigma_t=\Sigma_{t-1}+ \frac{1}{t}((X_t-\mu_{t})(X_t-\mu_{t})^T-\Sigma_{t-1})
\end{equation}
\item Increment $t$.
\end{enumerate}

In practice, this algorithm requires developments in order to be used in non-trivial conditions.
If we try to start the adaptation from the very first steps, the correction we make to the initial guess matrix $\Sigma_0$ is too big and numerically unstable. We thus introduced the parameter $t_0$, that can be tuned, and that represents the number of steps we wait before starting the adaptation. We use as default $t_0=2000$ steps which gave us good results in all the cases we studied (from N=20 to 50 parameters). If you are pretty sure of your initial matrix you may lower this number. Anyway the algorithm is very weakly dependent on this choice.
In the $t \leq t_0$ phase the chain starts exploring the parameter space, and it is not essential to have a very reliable first estimate for the covariance matrix; it is preferable to have a high acceptance rate and underestimate the optimal step than to risk remaining stuck at the starting point. The scale $c$ in the proposal is thus chosen to be quite small (default 0.001)

Then at step $t_0$ we calculate the sample variance of the chain. The estimate of the covariance between the parameter $j$ and the parameter $k$ is simply 
\begin{equation}
\Sigma_{jk}=\frac{1}{t_0-1}\sum_{i=1}^{t_0}(X_{ij}-\bar{X}_j)(X_{ik}-\bar{X}_k).
\end{equation}

This reconstructed matrix is a fair first guess of the covariance of the posterior. Hence, the scale $c$ is now set to the optimal value (for Gaussians) of $2.4^2/d$ where $d$ is the number of parameters \citep{Gelman_1996,Dunkley_2005}
and we start running the matrix adaptation.

Since the likelihood is not necessarily Gaussian and the matrix not perfect (otherwise we won't need to run!), we introduced another refinement that concerns the scale parameter $c$.
At a time $t_s$ we begin to adapt the scaling parameter according to
\begin{equation}
\label{eq:scale_ad}
c_{t+1}=c_t \cdot (1-\frac{1}{t}) + \frac{1}{t} (\mathrm{a.r.}-0.25)
\end{equation}
where "a.r" is the acceptance rate (ie. the number of accepted moves over the total) and is computed with the previous 100 samples. The purpose of this stage is to tune the scale parameter $c$ in order to reach an a.r of about 0.25, a classical "rule of thumb" for the Metropolis algorithm, and that gave indeed in our experience the best results.
The $t_s$ default in \camel is 10000 samples. You may change this number (with counts steps after $t_0$) but note that the $\frac{1}{t}$ factors in Eq.~\ref{eq:scale_ad} makes the scale factor converge rapidly to a constant and we end up with a classical Metropolis-Hastings algorithm.

Keeping memory of all its history, the process is no more Markovian. So, does the adaptive algorithm have the correct ergodicity properties? It is indeed mathematically difficult to deal with non basic MCMC methods, and a lot of the algorithm used by the scientific community are not assured to have the right properties for convergence. 
However in the Adaptative-Metropolis case, the convergence was proven \citep{Andrieu_2006} because 
the asymptotic dependency between the elements of the chain is weak enough to apply large number theorems.

\subsubsection{Runing AM}

Let us see how to use the AM.
One needs to specify a few extra information to the parameter file. Most have correct default values but you should at least define the number of samples ("length")  and the path to the initial proposal covariance matrix ("proposal\_cov"):
\begin{lstlisting}
length=100000  # specify the chain length
bunchSize=1000 # the chain is dumped every bunchSize
algo=ada       # default algorithm (or "metro" for standard Metropolis)
proposal_cov=/path/to/your/covmatrix.dat # path to your first guess covariance matrix
t0=2000        # number of steps for initial cov construction
ts=10000       # number of samples over t0 before starting the scale adapation
scale=0.001    # initials scaling (shoud be low)
#seed=12345     # seed for the random number generator for proposal: if not defined drawn randomly
#seedU=78776    # seed for uniform shots in algorithm: if not defined drawn randomly
\end{lstlisting}

Some important remarks
\begin{itemize}
\item the program will dump the data into a specified text file every \verb|bunchSize| iterations. This is interesting for running remotely (as in batches) and flushing the result through the network without too intensive IOs. You can then investigate results without waiting the end of the run.
\item there are 2 seeds for the random number generators. \textbf{If you fix these numbers you will recover the same results} which is probably not what you want if you run multiple chains (but is interesting for reproducibility). You can then implement some logic to draw some random numbers for these seeds, or do not specify them at all: then they will be drawn randomly from the machine environment.
\item the covariance matrix should be given as a text file. You can (should?) use the Hessian matrix of the minimization (see Sect.\ref{sec:hessian}). If not specified a diagonal matrix will be built from the parameter file "errors" but it is probably a bad idea to do so... 
\item because of the hopping nature of the algorithm, it is unnecessary (and CPU consuming) to use a high  precision file. Don't put any \verb|precisionFile| line in your parameter file: this will use \class default settings which are sufficient.
\end{itemize}

If you are using a full Boltzmann solver (as \class) you may also want to add to the output some \textit{derived} parameters which are computed at each sampled step. The list of them can found in the \verb|src/camel/Engine.hh::get| method. You add to your parameter file some lines as:
\begin{lstlisting}
derived=H0
derived=z_reio
derived=YHe
derived=sigma8
do_mPK=true
\end{lstlisting}
(note that you need to activate the "do\_mPk" flag for sigma8 (which turns on the matter power spectrum computations)

You may also add some redshift dependent values as 
\begin{lstlisting}
derived=sigma8(0.57)
derived=f(0.57)
class z_pk 0,0.57
\end{lstlisting}
The last line is here to ask the exact computation of P(k) at z=0 and 0.57.

In order to use a script to submit jobs with AM, just put the parameters discussed in this section in some text file (for instance \texttt{mcsetup.txt} and run:
\begin{verbatim}
runMCMC.sh /path/to/parfile mcsetup.txt
\end{verbatim}
This will launch 4 openMP jobs (you can change this number by redefining the NCHAINS environment variable).
The chains will be written as plain ASCII files under the name \verb|samples*.txt| in the directory created.

\subsubsection{Convergence}
We discuss here the question of MCMC convergence. We illustrate it with the chains produced with the 
\parfile run and with the settings discussed in the previous part.
Note that we used directly for the initial covariance matrix the Hessian obtained in Sect \ref{sec:mini}. We ran 4 chains of 150000 samples each which produced 4 output files.

There exists no exact mathematical way to determine when (if) the chain reached "convergence", ie. entered a regime where it samples correctly the likelihood (in all dimensions). But there are some tricks and we discuss here two classical ones.

The first thing you should look at is simply each variable evolution ("trace plots").
This is shown below for one chain (obviously all should be looked at).

\begin{figure}[h!]
\begin{center}
\includegraphics[width=.8\columnwidth]{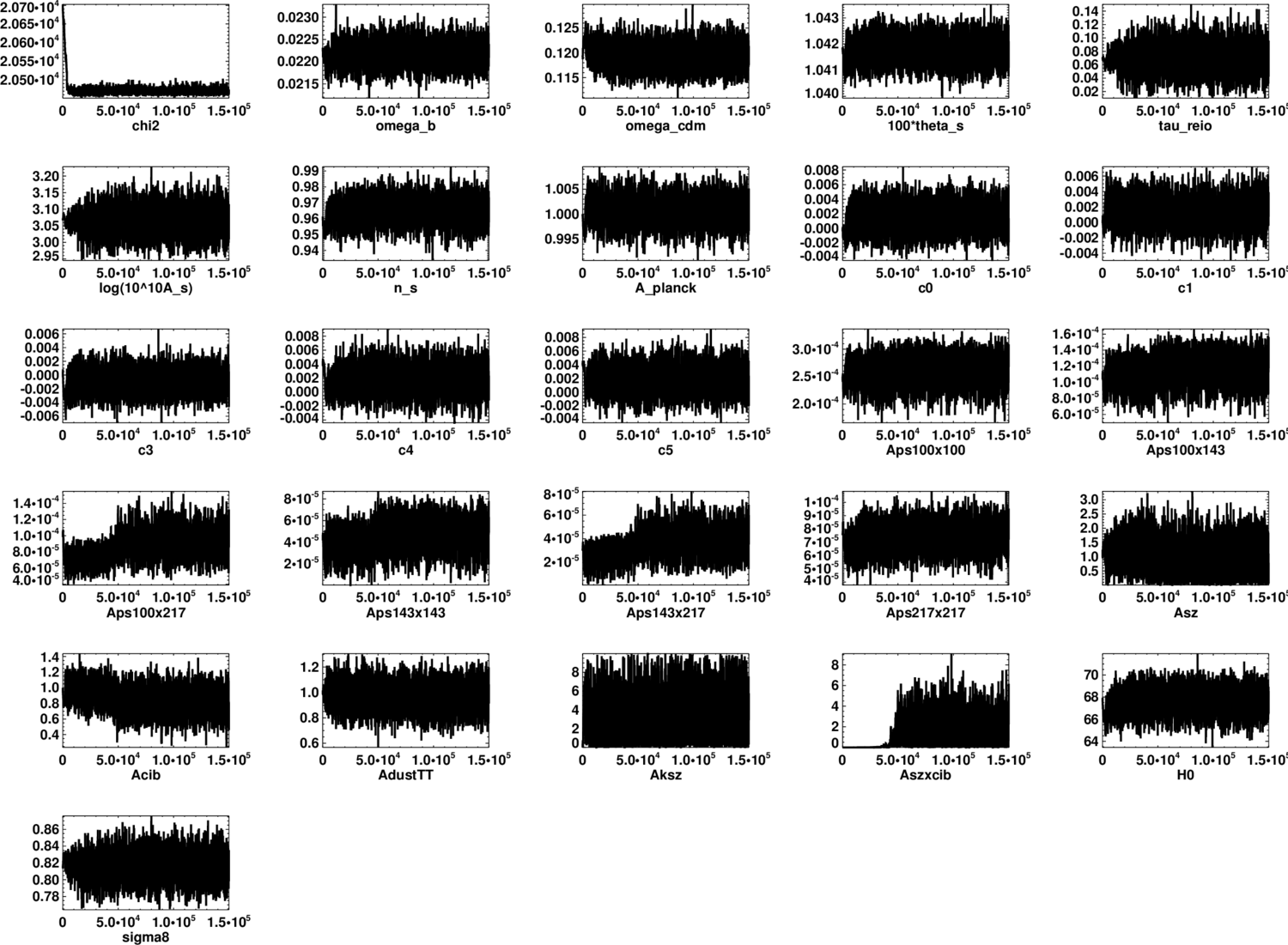}
\caption{{\label{fig:trace}
Plots of the values of each variable generated with the MCMC algorithm for one chain using \parfile. The first one displays the $\chi^2$ evolution. The last 2 variables are derived parameters.%
}}
\end{center}
\end{figure}

In this exercise for example, the "Aszxcib" variable began to be widely explored only lately (after $\simeq$50000 samples). This impacted the "Acib and all the "Aps" variables evolution.

This kind of plot only tests the \textit{stationarity} of the chains.
The real challenge however is to reach \textit{ergodicity} (which requires stationarity) meaning that the samples can be used to compute statistical expectation values (ie. "time" integrals can be used to replace "ensemble" averages)
This is the idea behind the Gelman-Rubin (GR )test \cite{Gelman_1992} which is run on a few chains and compares the inter- and cross- variance of the samples. It produces $R_i$ ,the "potential scale reduction factor" for each variable and the usual prescription is to assume that the variables reached ergodicity when all $R_i-1 \le 0.03$ (but you may feel more comfortable with 0.01).

\begin{figure}[h!]
\begin{center}
\includegraphics[width=0.7\columnwidth]{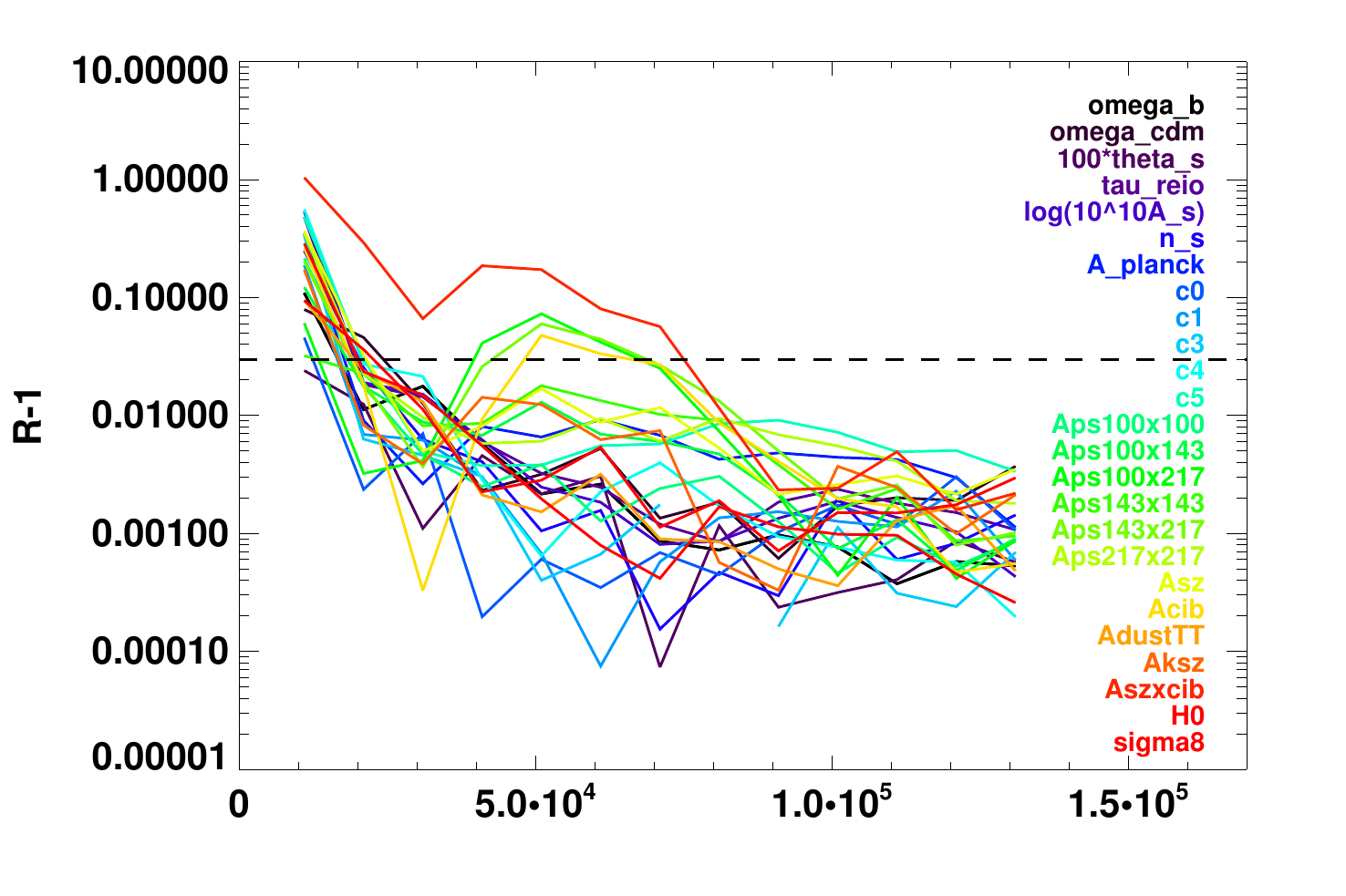}
\caption{{\label{fig:GR}
Gelman-Rubin test on the 4 chains for the sampled \parfile file. The standard cut is to use samples above $(R-1) \ge 0.03$ and is shown as a dashed line.%
}}
\end{center}
\end{figure}

The result in our example, as implemented in \camel in \texttt{python} or \texttt{idl}, is shown on Fig. \ref{fig:GR}.

The GR-test confirms that "Aszxcib" began to be sampled correctly only after $\gtrsim 80000$ steps. Once its sampling begun ($\gtrsim 50000$) it indeed affected the "Aps" and "Acib" variables which is reflected in some increase of their R values before finally converging to low values.

While a number of other tests exists (we also implemented the Geweke one), GR produces generally satisfactory results (for simple posteriors).

\subsubsection{MCMC analysis}
Given the GR-test, we decided to keep samples above 100000 steps and concatenate them into a single "well-converged" chain.

\begin{figure}[h!]
\begin{center}
\includegraphics[width=1\columnwidth]{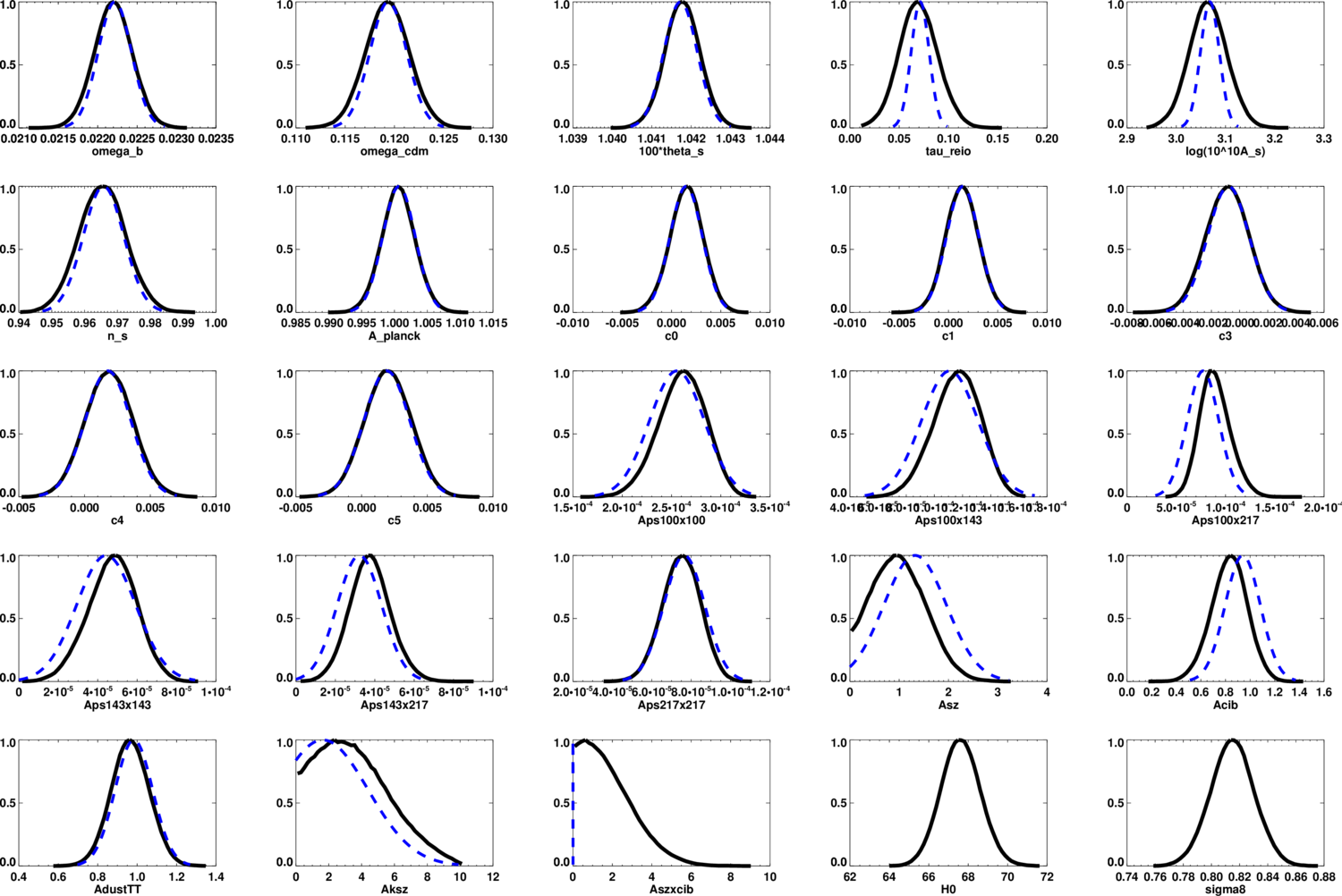}
\caption{{\label{fig:post1}
full lines: 1D posterior distributions of MCMC samples obtained by histograming + smoothing the well-converged chain. For clarity the maximum is normalized to 1. We also superimpose in dashed the output of the best-fit values with errors taken from the Hessian matrix. The last 2 parameters are MCMC-derived ones and have thus no associated best-fits.%
}}
\end{center}
\end{figure}

1D posterior distributions are then obtained by histograming (and possibly smoothing) each variable individually .
They are shown on Fig. \ref{fig:post1}. For further discussion, we also superimpose the best-fit values with Gaussian errors taken from the diagonal of the Hessian matrix.

First we note that the best-fit positions (dashed curve peaks) sometimes do not fall on the maximum of the posteriors (the "mode"). This is known as "volume effects" and will be discussed in Sect. \ref{sec:vol}.
Then the width of the distributions as inferred from the Hessian diagonal elements are quite good, but for "AszxCIB" which was stuck at 0 during the minimization.
This explains why it took some time for this parameter to be well sampled as was singled out by the GR-test. Despite this very wrong input, the algorithm still converged in a reasonable time.

\begin{figure}[h!]
\begin{center}
\includegraphics[width=1\columnwidth]{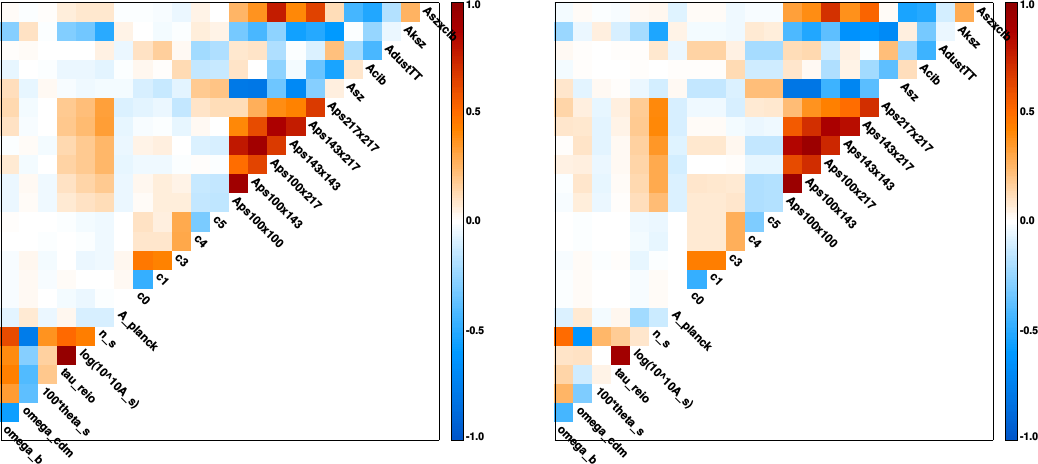}
\caption{{\label{fig:cors}
Correlation matrix between all the parameters estimated from the sampled posterior (on the left) and the Hessian matrix (on the right).%
}}
\end{center}
\end{figure}

Let us now study the correlations among the parameters.
One can compute the empirical covariance matrix between the estimated parameters
\begin{align}
\hat C=\dfrac{1}{N-1} \sum_{i=1}^N \Delta_i^{T} \Delta_i
\end{align}
where the $i^{th}$ realization deviation vector is  $\Delta_i \equiv \vec{p}_i-\hat \mu$,  $\vec{p}$ being the vector of parameters and $\hat \mu$ its statistical mean.
We convert it to a correlation matrix and show it on the left of Fig. \ref{fig:cors} (we drop the "H0" and "sigma8" parameters which are derived parameters).

The comparison with the Hessian matrix on Fig. \ref{fig:cors}(b) reveals a posteriori that \minuit performed quite well in determining the correlation structure. However note that the bottom part (cosmological parameters) is not perfect.

\begin{figure}[h!]
\begin{center}
\includegraphics[width=0.8\columnwidth]{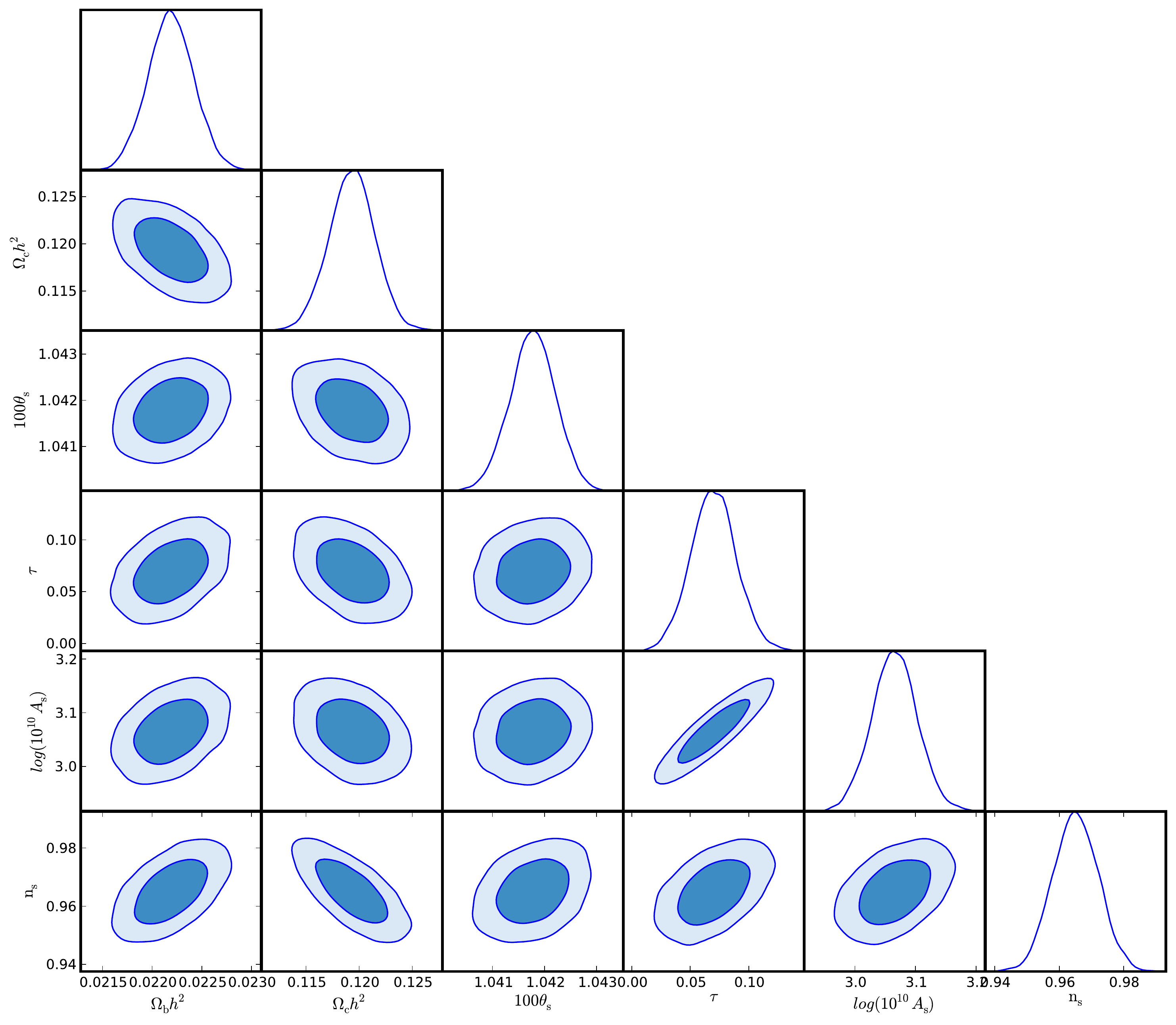}
\caption{{\label{fig:triangle}
"Triangle plot" for the MCMC samples showing the 1D and 2D posterior distributions of the cosmological parameters from the \parfile run. Ellipses show the 68 and 95\% credible intervals.%
}}
\end{center}
\end{figure}

A famous representation in a single figure of both the 1D and 2D posterior distributions is in the form of a "triangle plot" available in \camel through the \verb|camel.py| library or through the \texttt{GetDist} package (that can read \camel's chains) and is shown on Fig. \ref{fig:triangle}.

\pagebreak
\section{Comparison of the results}
\label{sec:comp}
\subsection{\lcdm\ parameters}

\begin{figure}[h!]
\begin{center}
\includegraphics[width=0.7\columnwidth]{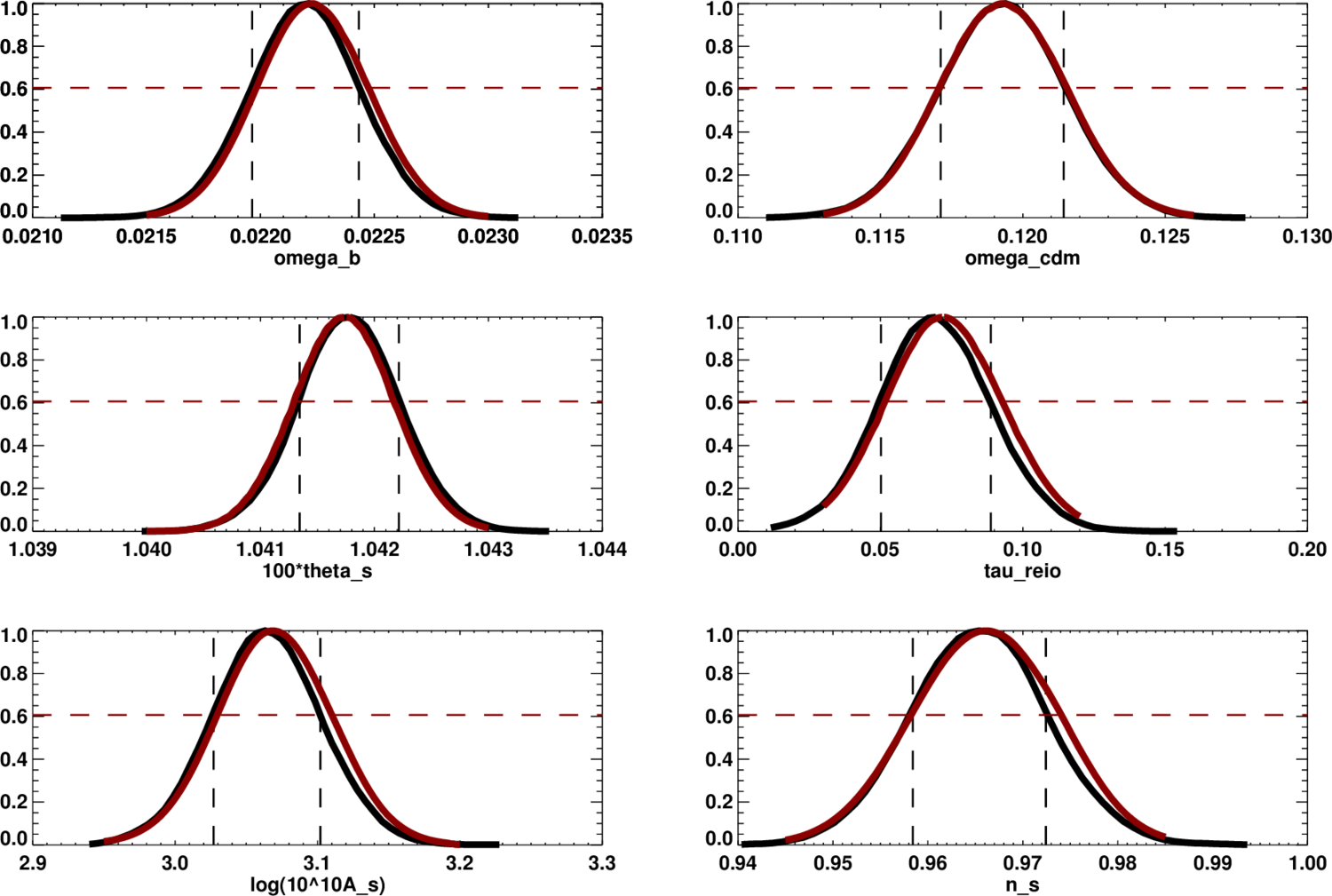}
\caption{{\label{fig:prof_vs_mc}
The MCMC posterior distributions (in black) as shown previously on Fig. \ref{fig:post1} are compared to the profile-likelihoods (in red , same than those on Fig. \ref{fig:lcdm_prof} but represented here as  $e^{-\dchi/2}$) for the \lcdm\ cosmological parameters using in both cases the same \parfile setup. The dashed lines are here to recall that 68\% level interval are not obtained in the same way: for posteriors the interval is given on the abscissa (as mean $\pm 1 \sigma$), while for profile likelihood the prescription $\dchi=1 \implies \lik/\lik_\textrm{max}= e^{-1/2}$ defines the cut along the ordinate.  Both intervals agree in the Gaussian case.%
}}
\end{center}
\end{figure}

We can now check if \textit{in this working case} both statistical methodologies lead to the same intervals.
We show on Fig \ref{fig:prof_vs_mc} the comparison between the 1D posteriors and the profile-likelihoods of the cosmological parameters as obtained in the previous sections. For this latter and the sake comparison, we represent $e^{-\dchi/2}$ and the $\dchi=1$ cut translates into a $e^{-1/2}\simeq 0.6$ one.

We therefore see that for \lcdm using the \planck\ \hlp+\bflike\ likelihoods both the Bayesian and frequentist approaches give very similar results on cosmological parameters, a result coarsely shown with other \planck\ likelihoods \citep{prof14,cosmo15}.

One should not conclude however it is a general statement: it happens here because the data constrain
well all the parameters, in particular the use of the \lowell\ likelihood breaks the $(\tau,\lnAs)$ degeneracy from the \hiell\ part.
The posteriors are close to Gaussian so that their $1\sigma$ interval corresponds neatly to the one obtained with the profile-likelihoods cut at $e^{-1/2}$. More generally, for a single parameter, the Bayesian credible interval will be a genuine "confidence level" if and only if the parameter is a "location" one, ie. $\lik(x;\theta)=\lik(x-\theta) $ or some transformation of the likelihood leads to it \citep{Porter96}. In the multi-dimensional case, some differences may appear due to the Bayesian marginalization. This is called the "volume effect" and is discussed next. 

\subsection{Likelihood volume effects}
\label{sec:vol}
\begin{figure}[h!]
 \begin{center}
 \includegraphics[width=0.3\columnwidth]{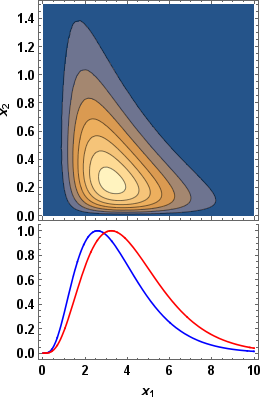}
 \caption{{\label{fig:toy}
 A toy example of a 2D likelihood (upper plot) and (bottom plots)the corresponding distributions for the $x_1$ marginalized posterior (in blue) and its profile-likelihood (in red).%
 }}
 \end{center}
 \end{figure}

In order to illustrate the difference between the profile-likelihoods and posterior distributions with several parameters, we begin with a toy example. Suppose the likelihood has 2 parameters ($x_1,x_2)$ and that its shape is of the type shown on the upper plot of Fig. \ref{fig:toy}. Then the marginalized posterior for $x_1$ reads
\begin{align}
p(x_1)\propto\int \lik(x_1,x_2) dx_2,
\end{align}
while the profile-likelihood is
\begin{align}
\lik(x_1)\propto \max_{x_2} \lik(x_1,x_2).
\end{align}
One can then easily understand why the maximum of the profile-likelihood lies at the same position than the 2D maximum.
This is not necessarily the case for the marginalized posterior where some part of the "volume" of the 2D likelihood may shadow the projection. Indeed these methods answer different questions: in the case of the posterior "what is the value of $x_1$ given the possible $x_2$ values?", while for profile-likelihood, "what is the value $x_1$ for the most likely $x_2$ values?".

We now give an example where this situation happens.
It was shown in \citet{papierAlens} that when removing the \lowell\ part of the \planck\ baseline likelihoods and keeping only the \hiell\ one (named \texttt{Plik}), the profile-likelihood method leads to an optical reionization depth of:
\begin{align}
\label{eq:profHFI}
\hat \tau=0.17\pm0.04 ~\onesig{Planck-\hiell/profile-likelihood}.
\end{align}

\begin{figure}[h!]
 \begin{center}
 \includegraphics[width=0.5\columnwidth]{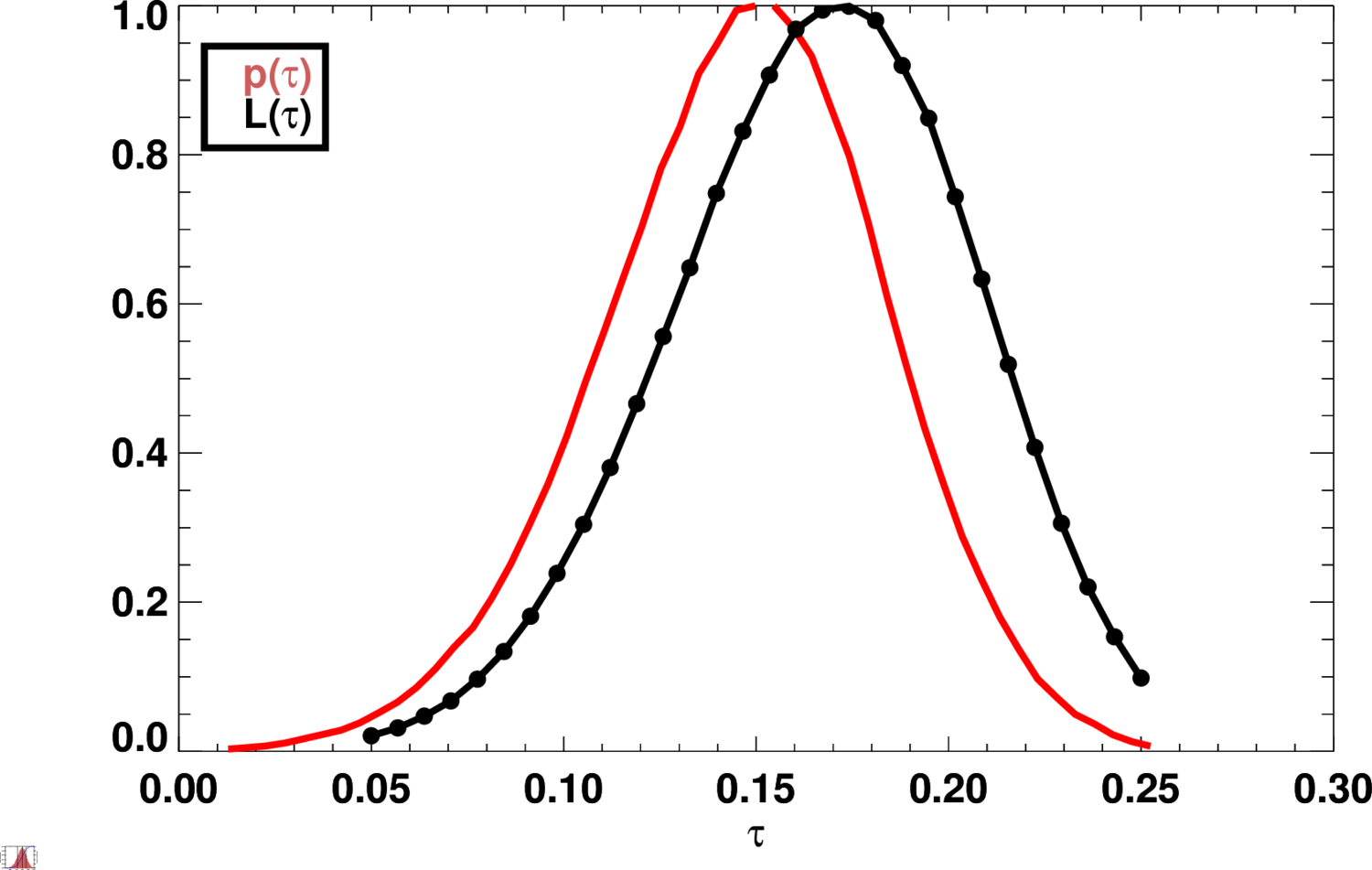}
 \caption{{\label{fig:likvspost}
 Reionization optical depth posterior (in red) obtained by sampling, with the AM algorithm, the \planck\ \hiell-only likelihood ({\tt plik}). The corresponding profile likelihood is superimposed in black.%
 }}
 \end{center}
 \end{figure}

The $\tau$ posterior distribution, obtained by running the AM method, is shifted to lower values is as shown on Fig.\ref{fig:likvspost}.
The peak of the distribution (the "mode") is located around 0.15 , \footnote{and even slightly less for the mean and median}.
By integrating the tails, one can reconstruct the 68\% central credible interval, which gives $[0.11,0.18]$ and taking the mode as the best estimate, we obtain:
\begin{align}
\label{eq:credible}
\hat \tau=0.15^{+0.03}_{-0.04} ~\onesig{Planck-\hiell/posterior}.
\end{align}

The choice between both methodologies can play a role.
For instance, from an LFI-based \lowell\ analysis \citep{Like15}, \planck\ reports a value of \footnote{ This result is independent of the methodology used}: $\hat \tau=0.067^{+0.023}_{-0.021} \onesig{Planck-LFI \lowell}$.
It is in $2.2\sigma$ tension with Eq.\ref{eq:profHFI} but $1.8\sigma$ with Eq.\ref{eq:credible}.
Not over-emphasizing the $2\sigma$ boundary, the use of the Bayesian result may have hidden this tension that is now increased 
by the new HFI-based result \citep{reio}: $\hat \tau=0.058\pm0.012,\onesig{Planck-HFI \lowell}$ to respectively $2.6\sigma$ (profile) and $2.2\sigma$ (posterior). 

\section{Summary}
We have seen how comparing different statistical approaches can be fruitful in cosmological analyzes.
A possible workflow can be:
\begin{enumerate}
\item first run an accurate Minimization with \minuit with a few starting points. Obtain the best-fit and the Hessian matrix.
\item use the Hessian as the starting matrix to the proposal of the Adaptive Metroplis MCMC algorithm.
\item if some posterior modes differ significantly from the best-fit value, run single profile-likelihoods on those parameters.
\item if some parameters are of particular importance (as $(w_0,w_a)$ in future surveys \footnote{2D profile-likelihoods are also implemented}), always run a profile-likelihood on them to verify the influence of priors and volume effects in the MCMC run.
\end{enumerate}
\camel provides the tools to perform such rich studies \textit{using the very same input file}. See the \url{camel.inp3.fr} web site for more information. The software is collaborative and contributions are most welcome.

\bibliography{bibliography/biblio}

\pagebreak
\appendix
\section*{\label{sec:hl_phuge} Appendix A : a (very) complete parameter file}
The following file shows the complete list of likelihoods with their associated nuisance parameters. The Engine used is \class.
\begin{lstlisting}{caption=The CAMEL master parameter file}
##############################################################
#Boltzman Engine
engine=class

#class setup
class     N_ncdm                1
class     m_ncdm                0.06
class     N_eff                 2.046

class     k_pivot               0.05
class     lensing               yes
class     sBBN\ file            bbn/sBBN.dat

precisionFile=class_pre/hpjul2.pre
###############################################################



###############################################################
#Cosmological model
par   omega_b             cosm    0.02224      0.00027   0.017   0.027
par   omega_cdm           cosm    0.1192      0.0026    0.09    0.15
par   100*theta_s         cosm    1.0418      0.6E-04   1.03	1.05
par   tau_reio            cosm    0.07	      0.13E-01  0.01	0.20
par   log(10^10A_s)       cosm    3.07	      0.025     2.7     3.5
par   n_s                 cosm    0.96	      0.0070    0.9     1.1
###############################################################



###############################################################
#LIKELIHOODS

#-----------------------------
#CMB low-l (BFLIKE)
clikfile=planck_data/low_l/bflike/lowl_SMW_70_dx11d_2014_10_03_v5c_Ap.clik
#-----------------------------


#-----------------------------
#CMB high-l (HILLIPOP)
HiLLiPOP=HiLLiPOP/DX11dHM_superExt_CO_TT.lik
par   A_planck     	  nui	  1		0.001	     0.9 	1.1
par   c0		  nui     0.		0.001     -0.05   0.05
par   c1		  nui     0.		0.001     -0.05   0.05
fix   c2		  nui     0.		0.001     -0.05   0.05
par   c3		  nui     0.		0.001     -0.05   0.05
par   c4		  nui     0.004		0.001     -0.05   0.05
par   c5		  nui     0.004		0.001     -0.05   0.05
par   Aps100x100          nui     2.5E-04	1E-05     0.0     0.1
par   Aps100x143          nui     1.1E-04	7E-06     0.0     0.1
par   Aps100x217          nui     9.9E-05	6E-06     0.0     0.1
par   Aps143x143          nui     4.7E-05	2E-06     0.0     0.1
par   Aps143x217          nui     3.1E-05	3E-06     0.0     0.1
par   Aps217x217          nui     7.6E-05	6E-06     0.0     0.1
par   Asz                 nui     1		0.1       0.0     10
par   Acib                nui     1.		0.1       0.0     10
par   AdustTT		  nui	  1	        0.1	  0.0     2
fix   AdustPP		  nui	  0.00        0.1	  0.0     2
fix   AdustTP		  nui	  0.00        0.1	  0.0     2
par   Aksz                nui     0.00        1.0	  0.0     10
par   Aszxcib             nui     0.00        1.0	  0.0     10

#PRIORS
gauss1    AdustTT	1	0.2
gauss1    Acib		1	0.2
gauss1    A_planck	1	0.0025
gauss1    c0		0	2e-3
gauss1	  c1		0	2e-3
gauss1	  c3		0	2e-3 
gauss1	  c4		0.002	2e-3 
gauss1	  c5		0.002	2e-3 
#-----------------------------


#-----------------------------
#CMB Very-high-l (SPT)
SPT_High=HighEll/SPT_high_2014.lik
par     SPT_high_95_cal         nui     0.9961  0.002   0.9     1.1
par     SPT_high_150_cal        nui     1.002   0.002   0.9     1.1
par     SPT_high_220_cal        nui     1.015   0.002   0.9     1.1
par     SPT_high_Aps_95x95      nui     7.425   0.01    0.1     50
par     SPT_high_Aps_95x150     nui     5.147   0.01    0.1     50
par     SPT_high_Aps_95x220     nui     8.8     0.01    0.1     50
par     SPT_high_Aps_150x150    nui     6.649   0.01    0.2     50
par     SPT_high_Aps_150x220    nui     14.15   0.01    1.5     50
par     SPT_high_Aps_220x220    nui     36.07   0.01    3       200
fix     SPT_ADust               nui     1

#PRIORS
gauss1 SPT_high_95_cal 1.01 .01
gauss1 SPT_high_150_cal 1.01 .01
gauss1 SPT_high_220_cal 1.01 .02 
#-----------------------------


#-----------------------------
#CMB Very-high-l (SPT-low)
SPT_Low    = HighEll/SPT_low.lik
par             SPT_low_Aps             nui             20.32           0.01		1.              60
par             SPT_low_cal             nui             1.00		0.01            0               2            
#-----------------------------


#-----------------------------
#CMB Very-high-l (ACT)
ACT_equat=HighEll/ACT_equat.lik
par     ACT_equat_148_cal       nui     0.9991  0.002   0.9     1.1
par     ACT_equat_220_cal       nui     1.013   0.002   0.9     1.1
par     ACT_equat_ADust         nui     1.719   0.01    0.05    10
par     ACT_equat_Aps_148x148   nui     7.159   0.01    0.1     50
par     ACT_equat_Aps_148x220   nui     20      0.01    0.1     50
par     ACT_equat_Aps_220x220   nui     60      0.01    10      150
ACT_south=HighEll/ACT_south.lik
par     ACT_south_148_cal       nui     1.007   0.002   0.9     1.1
par     ACT_south_220_cal       nui     1.032   0.002   0.9     1.1
par     ACT_south_ADust         nui     1.3     0.01    0.05    10
par     ACT_south_Aps_148x148   nui     9       0.01    0.1     50
par     ACT_south_Aps_148x220   nui     16.29   0.01    0.1     50
par     ACT_south_Aps_220x220   nui     60      0.01    10      150
#-----------------------------


#-----------------------------
#CMB lensing (Planck)
lensing_file=planck_data/lensing/smica_g30_ftl_full_pttptt.clik_lensing#
#-----------------------------


#-----------------------------
#SNIa (JLA)
JLA_SNIA_File=JLA/JLA.lik
par       Mabs          nui     -19.04          .03             -19.25          -18.85
par       alpha         nui     0.141           .01             0.10            .18
par       beta          nui     3.101           0.1             2.5             3.7
par       DeltaM	nui     -.076           0.03            -.13            -0.01 
#-----------------------------


#-----------------------------
#BAO
BAOFile=BAO/BAO1D_new.lik
BAO2DFile=BAO/Boss_anderson_dr11.lik
#-----------------------------


###############################################################




###############################################################
#fitter
nitermax=50000
set_stra=2
set_tol=0.00001

#output options
remove_cosmo_limits=false
doHesse=true
###############################################################
\end{lstlisting}

\end{document}